\documentclass[ba]{imsart}
\pubyear{0000}
\volume{00}
\issue{0}
\doi{0000}
\firstpage{1}
\lastpage{1}

\usepackage{amsthm}
\usepackage{amsmath}
\usepackage[colorlinks,citecolor=blue,urlcolor=blue,filecolor=blue,backref=page]{hyperref}
\usepackage{graphicx}
\usepackage{caption}
\usepackage{subcaption}

\startlocaldefs
\numberwithin{equation}{section}
\theoremstyle{plain}

\newcommand\independent{\protect\mathpalette{\protect\independenT}{\perp}}
\def\independenT#1#2{\mathrel{\rlap{$#1#2$}\mkern2mu{#1#2}}}

\endlocaldefs

\newcommand{\expe}[1]{\operatorname{\mathbb{E}}\left(#1\right)}

\newcommand{\bsl}{\begin{slide}}
\newcommand{\esl}{\end{slide}}

\newcommand{\bm}{\mathbf{m}}

\newcommand{\sgn}{\mbox{sgn} }

\newcommand{\bei}{\begin{itemize}}
\newcommand{\eni}{\end{itemize}}

\newcommand{\beq}{\begin{equation}}
\newcommand{\bdm}{\begin{displaymath}}
\newcommand{\edm}{\end{displaymath}}

\usepackage{times}
\usepackage{bm}

\usepackage[plain,noend]{algorithm2e}

\makeatletter
\renewcommand{\algocf@captiontext}[2]{#1\algocf@typo. \AlCapFnt{}#2} 
\renewcommand{\AlTitleFnt}[1]{#1\unskip}
\def\@algocf@capt@plain{top}
\renewcommand{\algocf@makecaption}[2]{%
	\addtolength{\hsize}{\algomargin}%
	\sbox\@tempboxa{\algocf@captiontext{#1}{#2}}%
	\ifdim\wd\@tempboxa >\hsize
	\hskip .5\algomargin%
	\parbox[t]{\hsize}{\algocf@captiontext{#1}{#2}}
	\else%
	\global\@minipagefalse%
	\hbox to\hsize{\box\@tempboxa}
	\fi%
	\addtolength{\hsize}{-\algomargin}%
}
\makeatother

\usepackage{amssymb,lscape,psfrag}
\usepackage{epsfig,color}
\usepackage{array}
\usepackage{enumerate}
\usepackage{amsmath, amsthm}
\numberwithin{equation}{section}
\usepackage{graphics}
\usepackage{graphicx}
\usepackage{epsfig}
\usepackage{verbatim}
\usepackage{cancel}
\usepackage{longtable}
\usepackage{multirow}
\usepackage{float}
\usepackage{ulem}
\usepackage[toc,page]{appendix}
\usepackage{rotating}
\usepackage{amsmath}
\usepackage{url}

\usepackage{verbatim}

\usepackage[comma,longnamesfirst]{natbib}

\newcommand {\ee} {\end {equation} }
\newcommand {\bmath} {\begin {displaymath} }
\newcommand {\emath} {\end {displaymath} }

\newcommand {\diag} { \mbox{diag} }

\begin{document}
	\begin{center}
		
		{\Large  \bf Approximate Bayesian Conditional Copulas}
		
		\medskip
		
		{\em  Clara Grazian, Luciana Dalla Valle, Brunero Liseo}
		
		{\textit{University of New South Wales, ACEMS, University of Plymouth and Sapienza Universit\'a di Roma}}
		
	\end{center}
	
	\hrulefill\

	\vspace{.41cm}

\section*{Abstract}	

Copula models are flexible tools to represent complex structures of dependence for multivariate random variables. 
According to Sklar's theorem \citep{sklar1959fonctions}, any $d$-dimensional absolutely continuous density can be uniquely represented as the product of the marginal distributions and a copula function which captures the dependence structure 
among the vector components.
In real data applications, the interest of the analyses often lies on specific functionals of the dependence, which quantify aspects of it in a few numerical values. A broad literature exists on such functionals, however extensions to include covariates are still limited. This is mainly due to the lack of unbiased estimators of the copula function, especially when one does not have enough information to select the copula model. 
Recent advances in computational methodologies and algorithms have allowed inference in the presence of complicated likelihood functions, especially in the Bayesian approach, whose methods, despite being computationally intensive, allow us to better evaluate the uncertainty of the estimates.
In this work, we present several Bayesian methods to approximate the posterior distribution of functionals of the dependence, using nonparametric models which avoid the selection of the copula function. 
These methods are compared in simulation studies and in two realistic applications, from civil engineering and astrophysics. 

\vspace{0.2cm}

\textbf{Keywords:} approximate Bayesian computation, conditional copulas, Bayesian inference, dependence modelling, Gaussian processes, empirical likelihood, splines.

\section{Introduction}	
\label{sec:intro}

Copula models have received an increasing interest since the work of \cite{sklar1959fonctions}.
Sklar's theorem is a probability result which states that every multivariate cumulative distribution function (CDF, hereafter)
can be represented as the product of the marginal distributions times a factor, namely the copula function, which captures the dependence structure among the components of the random vector.
This result is very important in statistical modelling, especially when it is reasonable and useful to separately model the marginal distributions and the potentially complex multivariate dependence structure, or when the degree of information about the marginals and their dependencies is different, since, in general, more information can be gathered on marginal aspects of the problem at hand.

Sklar's theorem proves that every multivariate distribution function $F_{\bm{Y}}(\cdot)$ of a random variable $\mathbf{Y} = (Y_1,\dots,Y_d)$ can be represented by a copula function $C(\cdot): [0,1]^d \rightarrow [0,1]$ depending on $d$ univariate marginal distributions
\begin{equation}
F_{\bm{Y}}(y_1,\dots, y_d ) = C\left (F_1(y_1|\boldsymbol{\theta}_1), \dots, F_d(y_d|\boldsymbol{\theta}_d) \mid  \boldsymbol{\psi}\right) 
\label{eq:uncond_joint}
\end{equation}
where $F_j(\cdot)$ is the marginal CDF of $Y_j$, depending on the parameter vector $\boldsymbol{\theta}_j$, for $j=1,\dots,d$, and $\boldsymbol{\psi}$ is the copula parameter vector. This representation is unique if 
$F_{\bm{Y}}$ is continuous. 
In the case of continuous random variables, Equation \eqref{eq:uncond_joint} admits the following density, which shows how the copula function absorbs all the dependence of the model  
\begin{equation*}
f_{\bm{Y}}(y_1,\dots,y_d) = c(F_1(y_1|\boldsymbol{\theta}_1), \dots, F_d(y_d|\boldsymbol{\theta}_d) \mid \boldsymbol{\psi}) \cdot f_1(y_1|\boldsymbol{\theta}_1) \cdot \ldots \cdot  f_d(y_d|\boldsymbol{\theta}_d)
\end{equation*}
where $c(\cdot)$ is the density of the copula function $C(\cdot)$.
We refer the reader to \cite{nelsen2007introduction} for a detailed description of copula theory and methods. 

\cite{patton2006modelling} extends the definition of copula in the presence of covariates, to describe situations where the marginals and their dependence structure are influenced by the values of other variables, that is:
\begin{equation*}
F_{\bm{Y}|X}(y_1,\dots, y_d | X=x) = C_X \left ( F_{1| X}(y_1 | X=x), \dots,  F_{d | X}(y_d | X=x) \, | \, X=x
\right ) 
\end{equation*}
where $X \in \mathbb{R}^p$ represents a set of covariates, 
$C_X(\cdot)$ is the conditional copula, which may potentially vary with $X$, and 
$F_{j | X}$ is the conditional CDF of $Y_j$, for $j=1,\dots, d$. Here and in the following, the dependence on the parameters $(\boldsymbol{\psi}, \boldsymbol{\theta}_1, \dots, \boldsymbol{\theta}_d)$ is left implicit in the notation of the CDFs. 
The introduction of covariates may be useful in many applications where the dependence structure varies over the space of the observations \citep{acar2011dependence}.  
Moreover, conditional copulas are the building blocks of vine copulas \citep{czado2010pair}, where situations of dependence among the variables on which copulas are conditioned on are  common in real applications; in these cases, a  ``Simplifying Assumption'' \citep{czado2019analyzing} is often introduced in order to make statistical analysis easier. 
According to the Simplifying Assumption, the conditional copula is assumed constant, as in \cite{gijbels2015estimation}. Several contributions to the literature aim at exploring and testing violations of this assumption, such as in \cite{haff2010simplified}, \cite{acar2012beyond}, \cite{acar2013statistical}, \cite{killiches2016using}, \cite{killiches2017examination} and \cite{kurz2017testing}. 
However, \cite{levi2018bayesian} show that violations of the Simplifying Assumption may be due to the omission of important covariates, rather than to a real dependence on the included covariates. This result suggests that, in practical situations, it is safer to assume the potential dependence of the copula on the values of the available covariates.

A standard approach to model the influence of covariates on copulas is based on a parametric model which assumes a functional relationship between copula parameters and covariates, such that 
$C_X(\cdot) = C_{\psi(X)}(\cdot)$, 
where $\psi \in \Psi$ is the copula parameter, assumed to be a function of the covariates $X$, supposing for simplicity that the copula depends on one parameter. In this setting, the parameter is associated to the covariates through a link function $\zeta: \Psi \rightarrow \mathbb{R}$, such that $\psi(X) = \zeta^{-1}(\eta(X))$, where $\eta(\cdot)$ is a real-valued calibration function.  
The calibration function may assume different forms. 
A parametric form is adopted, for example, by \cite{genest1995semiparametric}, while a nonparametric form is suggested by \cite{acar2011dependence}, who employ a local polynomial-based approach, and \cite{craiu2012mixed}, \cite{vatter2015generalized}, \cite{klein2016simultaneous} and \cite{stander2019analysis}, who propose additive conditional copula regression specifications with predictors defined using splines. 

Different approaches are considered in the literature for the estimation of conditional copulas. 
\cite{abegaz2012semiparametric} and \cite{gijbels2012multivariate} have proposed  semiparametric and nonparametric methodologies within the frequentist framework to model the influence on copulas of covariates
taking values in complex
spaces; in both papers the authors consider the statistical properties of conditional copula estimators, establishing
consistency and asymptotic normality results.

In the Bayesian framework, \cite{valle2018bayesian} have proposed to nonparametrically estimate the conditional copula density in the bivariate case, introducing a generalization of ideas presented in \cite{wu2015bayesian}. 
The authors assume that the unknown conditional copula density can be represented 
as an infinite mixture of Gaussian copulas, where the correlation parameter is defined as a (linear or non-linear) function of a covariate:
\begin{equation*}
c_X(u_1, u_2 \mid X=x) = \sum_{\iota=1}^{\infty} \pi_\iota c_{\rho(x)}(u_1, u_2 \mid X=x)
\end{equation*}
where $c_{\rho(x)}(\cdot)$ denotes the Gaussian copula densities with correlation coefficient $\rho(x)$ depending on the covariate $X$,
$u_1 = F_{1 \mid X}(y_1 \mid X=x)$, $u_2 = F_{2 \mid X}(y_2 \mid X=x)$, $\sum_{\iota=1}^{\infty} \pi_\iota = 1$ and $0 < \pi_\iota < 1$. 
The model can be described via hierarchical specification, for $i=1,\dots,n$, as
\begin{align*}
(u_{1i}, u_{2i}) \mid \rho(x_i \mid \boldsymbol{\beta}_i) & \sim c_{\rho(x_i \mid \boldsymbol{\beta}_i)}(u_{1i}, u_{2i} \mid x_i) \\
\boldsymbol{\beta}_i \mid G & \overset{\text{iid}}{\sim} G \\
G & \sim DP(\alpha, G_0).
\end{align*}
The coefficients $\boldsymbol{\beta}$, which relate the covariate to the correlation parameter $\rho$, are given a random CDF $G$ which follows a Dirichlet process prior with mass parameter $\alpha$ and base distribution $G_0$. The correlation parameter is then defined through a calibration function which is linear in the coefficients $\boldsymbol{\beta}$. 
The algorithm proposed by \cite{walker2007sampling} for mixture models is applied to implement the Bayesian sampling strategy. 

\cite{levi2018bayesian} propose to jointly estimate the marginal distributions and the copula function 
using Gaussian process (GP) models, where the calibration function follows \textit{a priori}
 a single-index model based on GP, to handle high-dimensional covariates. 
In more details, the authors assume that
$$
Y_j \sim \mathcal{N}(\boldsymbol{\eta}_j(x_i),\sigma^2_j \mathbf{I}_n), \hspace{1cm} j = 1,\dots,d, \hspace{1cm} i = 1,\dots,n,
$$
where $\mathcal{N}(\mu,\Sigma)$ is a multivariate Gaussian distribution with mean $\boldsymbol{\mu}$ and variance $\boldsymbol{\Sigma}$, $\boldsymbol{\eta}_j(x_i)$ is the calibration function for the marginal distribution $F_j$ and $\mathbf{I}_n$ is the identity matrix of dimension $n \times n$. 
The copula function is characterized by its own calibration function $\boldsymbol{\eta}(x_i) = \boldsymbol{\eta}(x_i^\prime \boldsymbol{\beta})$, for $i = 1,\dots,n$, where $\boldsymbol{\beta} \in \mathbb{R}^p$ is a set of coefficients that must be normalized for identifiability reasons ($|| \boldsymbol{\beta} || = 1$). Each calibration function follows a Gaussian process prior, centred around zero and with covariance structure depending on a kernel which is a function of the distance among covariates, e.g. the squared exponential kernel. 
Although the GP approach is very attractive for its flexibility,
the idea of modelling the parameters of a known copula as a function of covariates implies the need to choose the copula family. 
Several model selection methods, which may be applied to both the choice of the copula family and the choice of the form of the calibration function, are available. One approach compares the average prediction power of different models using the cross-validated pseudo-marginal likelihood (CVML) proposed by \cite{geisser1979predictive}; another approach is based on the Watanabe-Akaike information criterion, proposed by \cite{watanabe2013widely}. Both measures can be generalized to consider covariates. 

As a general comment, we point out that the choice of a statistical model for the distribution of a multivariate random vector is generally complex  and parametric assumptions are always difficult to verify. For this reason, it is often the case that the researcher prefers to reconsider the inference goals on some low dimensional functional of the copula, i.e., for example Kendall's $\tau$, Spearman's $\rho$ or some tail dependence indices; in such case, the complete dependence structure could be considered as a nuisance parameter.

\cite{grazian2017approximate} have derived an approximation of the posterior distribution $\pi(\varphi ; \mathbf{y})$ of the functional of interest $\varphi$ using 
\begin{equation*}
\pi(\varphi ; \mathbf{y}) \propto \pi(\varphi) \hat{L}(\varphi ; \mathbf{y})
\end{equation*}
where $\pi(\varphi)$ is a prior distribution and $\hat{L}(\cdot)$ is a nonparametric approximation of the likelihood function.  \cite{grazian2017approximate} use the exponentially tilted empirical likelihood proposed by \cite{schennach2005bayesian}. This version of the empirical likelihood allows for a Bayesian interpretation involving an implicit nonparametric process prior on the infinite-dimensional nuisance parameter (the copula structure).  Other version of the empirical likelihood \citep{owen2001empirical} can be used: see, for example, \cite{mengersen2013bayesian}.
This approach produces a good approximation of the posterior distribution (and of the likelihood function) if the generalized moment condition - which is implicit in the maximization problem associated with the definition of the empirical likelihood - is satisfied; this condition can be interpreted as a sort of unbiasedness requirement. In order to achieve this goal, a consistent estimator of the quantity of interest $\varphi$ is needed; while this is easy to do in an unconditional setting, common estimators of the conditional copula function -- upon which the estimators for the functionals are built -- have been shown to display some degree of inconsistency \citep{gijbels2011conditional}.

In this paper we explore several ways to make inference on functionals of the dependence in presence of covariates. The goal of the paper is to avoid the selection of the copula function. We discuss and propose three methods of increasing relaxation of the distributional assumptions on the functional of interest: 
the former is based on GPs, where, as opposed to \cite{levi2018bayesian}, the choice of a specific copula family is avoided;
the second is a direct generalization of the approach of \cite{grazian2017approximate} and makes use of the empirical likelihood of the functional of interest, either implementing an inconsistent estimator of the conditional copula or a linearized model; 
the latter makes use of Bayesian splines to approximate the behavior of the functional of the copula. 

The remainder of this paper is organised as follows: Section \ref{sec:method} presents several methods to perform inference on functionals of the dependence based on nonparametric representations of the copula function. Each of these methods is compared in Section \ref{sec:simu} and is contrasted to the conditional method proposed by \cite{levi2018bayesian} in the case of two covariates. Two real datasets showing non-linear dependence on one or two covariates are analysed in Section \ref{sec:app}, showing that areas of applications include a broad range spacing from civil engineering and energy management to astrophysics. Finally, Section \ref{sec:conclu} concludes the paper.

\section{Bayesian analysis for functionals of conditional copulas}
\label{sec:method}

\subsection{Conditional dependence measures}\label{sec:measures}

Let us consider the bivariate case and assume that for each level of a covariate $(x_1, x_2, \ldots, x_k)$ we observe $n_\ell$ replications of $Y_1$ and $Y_2$, with $\ell=1,\ldots,k$, and compute the probability integral transforms $u_j = F_{j \vert X_\ell}(y_j\vert X_\ell = x_\ell)$ for $j=1,2$ so to  obtain:
\begin{equation*}
\left[ \left( u_{1,1}, u_{2,1} \right), \ldots, \left( u_{1,n_\ell}, u_{2,n_\ell} \right) \right], \qquad \ell=1, \dots, k,
\end{equation*}
such that $n_\ell$ is the sample size of $(u_1,u_2)$ at location $x_\ell$ and $\sum_{\ell=1}^k n_\ell = n$.
The joint distribution is defined through a copula function:
\begin{equation*}
F_{Y_1, Y_2 | X}(u_1, u_2  \vert X=x) = C_X(u_1,u_2 \vert X=x).
\end{equation*}

We also assume that the strength of dependence between $U_1$ and $U_2$ can be modelled as a smooth function of $X$ and that we are not able to assume any specific parametric form for $C_X$. 
Also, we consider the situation where one is mainly interested in making inference on a synthetic dependence measure of $C_X$, say 
\begin{equation*}
\varphi(C_X; x) = \mathbb{E}[v(U_1,U_2)\big \vert  X=x].
\end{equation*}
for some function $v$. 
The quantities of interest are in general, functionals of the dependence, such as Kendall's $\tau$, Spearman's $\rho$ or tail dependence indices. Kendall's $\tau$ is a measure of similarity of the orderings of the data. 
Given two independent bivariate random variables $(Y_{1,1}, Y_{2,1})$ and 
$(Y_{1,2}, Y_{2,2})$, Kendall's $\tau$ is defined as 
$$
\tau = \expe{\sgn \left [ (Y_{1,1}- Y_{1,2})(Y_{2,1}-Y_{2,2}) \right ]}.
$$
This dependence measure can also be defined in terms of copulas as  
\begin{equation}
\tau(Y_1,Y_2) =  4 \int \int_{[0,1]^2} C(u_1,u_2) dC(u_1,u_2) - 1.
\label{eq:tau}
\end{equation}
The Spearman's $\rho$ is an alternative nonparametric measure of rank correlation, assessing how well the dependence among variables can be described by a monotonic function; it is defined as 
$$
\rho(Y_1,Y_2) = \textrm{Corr}\left ( F_{Y_1}(Y_1), F_{Y_2}(Y_2) \right);
$$
the corresponding copula expression is 
\begin{equation}
\rho(Y_1,Y_2) = 12 \int \int_{[0,1]^2} C(u_1,u_2) du_1 du_2 -3.
\label{eq:rho}
\end{equation} 
Other measures of interest are tail dependence indices or the conditional expected value of $U_1$ for a given level of $X$ and $U_2$, see \cite{levi2018bayesian}.


Since the above dependence indices can be directly defined through their copula function, the conditional versions of \eqref{eq:tau} and \eqref{eq:rho} can be easily derived in terms of conditional copulas:
\begin{equation}
\tau(x) =  4 \int \int_{[0,1]^2} C_X(u_1,u_2) dC_X(u_1,u_2) - 1, \qquad 
\rho(x) = 12 \int \int_{[0,1]^2} C_X(u_1,u_2) du_1 du_2 - 3.
\label{eq:cond_ind}
\end{equation}
Consequently, the most common estimators of measures of conditional dependence are expressed in terms 
of estimators of the conditional copula.
In particular, estimators of $\tau(x)$ and $\rho(x)$ are obtained in terms of
\begin{equation}
C_{X;h}(y_1,y_2) = \sum_{i=1}^n w_{n;i}(x,h_n) \mathbb{I}\left [Y_1 \leq y_1, Y_2 \leq y_2\right] 
\label{eq:estim_cond_cop}
\end{equation}
where $\mathbb{I}$ is an indicator function and $\{w_{n;i}(x, h_n)\}$ is a sequence of weights that smooth over the covariate space (for example, the Nadaraya-Watson or the local-linear weights) and $h_n>0$ is a bandwidth which is assumed to vanish as the sample size increases. See \cite{gijbels2011conditional} for a detailed definition. As a reminder, we report here the most common choices of weights. The Nadaraya-Watson weights are defined as
	\begin{equation*}
	w_{n;i}(x,h_n) = \frac{\mathcal{K}\left(\frac{X_i-x}{h_n}\right)}{\sum_\kappa\mathcal{K}\left(\frac{X_\kappa-x}{h_n}\right)},
	\end{equation*}
while the local-linear weights are defined as
	\begin{equation*}
	w_{n;i}(x,h_n) = \frac{\frac{1}{nh_n}\mathcal{K}\left(\frac{X_i-x}{h_n}\right)\left(S_{n,2} - \frac{X_i-x}{h_n}S_{n,1}\right)}{S_{n,0}S_{n,2}-S^2_{n,1}}	
	\end{equation*}
	where $S_{n,\kappa} = \frac{1}{nh_n} \sum_{i=1}^n \left( \frac{X_i - x}{h_n}\right)^\kappa \mathcal{K}\left(\frac{X_i - x}{h_n}\right)$. In both cases, it is necessary to choose a kernel smoothing over the covariate space: popular choices here are the triweight kernel
	\begin{equation*}
	\mathcal{K}(x) = \frac{35}{32} (1-x^2)^3 \mathbb{I}(|x|<1)
	\end{equation*}
and the Gaussian kernel
	\begin{equation*}
	\mathcal{K}(x) =\left( 2\pi\right)^{-1/2} \exp\left(-x^2/2 \right).
	\end{equation*}
These choices are the ones considered in this work. 
Based on these estimators of the copula function, \cite{gijbels2011conditional} propose the nonparametric version of the conditional Kendall's $\tau$:
\begin{equation*}
\hat{\tau}(x) = -1 + \frac{4}{1-\sum_{i=1}^n w^2_{n;i}(x,h_n)} \sum_{i=1}^n \sum_{t=1}^n w_{n;i}(x,h_n) w_{n;t}(x,h_n) \mathbb{I}(Y_{1i} < Y_{1t} , Y_{2i} < Y_{2t}), 
\end{equation*}
and of the conditional Spearman's $\rho$
\begin{equation*}
\hat{\rho}(x) = 12 \sum_{i=1}^n w_{n;i}(x,h_n) (1-\hat{U}_{1i})  (1-\hat{U}_{2i}) - 3, 
\end{equation*}
where $\hat{U}_{1i} = \hat{F}_{1|X}(y_{1i} | X_{i}=x_i)$ and $\hat{U}_{2i} = \hat{F}_{2|X}(y_{2i} | X_{i}=x_i)$.

Unfortunately, estimator \eqref{eq:estim_cond_cop} is biased and the size of the bias depends on the role of the covariates on the marginal distributions. In order to reduce the influence of the covariates, \cite{gijbels2011conditional} have proposed an alternative estimator, where the marginal distributions are similarly  approximated using  Equation \eqref{eq:estim_cond_cop}; in this case, a different sequence of weights must be adopted. 
Nonetheless, this latter estimator is biased as well: whereas, in particular settings, it can be drastically less biased than the former estimator, there is no guarantee that the bias will always be smaller than the bias of the former. In addition, it is necessary to choose three different bandwidth parameters in order to implement it. We refer to \cite{veraverbeke2011estimation} for a discussion of the asymptotic properties of these estimators. In conclusion, the problem of evaluating the bias of the nonparametric functionals of the dependence is still open. 

\cite{grazian2016approximate} have discussed a preliminary exploration of a potential extension of the approximate procedure described in \cite{grazian2017approximate}; 
they use estimators \eqref{eq:cond_ind} based on the empirical copula \eqref{eq:estim_cond_cop} and analyse their properties through simulation. 
However, the lack of theoretical warranties makes the performance of the procedure rather ``problem-specific'' and difficult to evaluate in general.

In this work, the goal of the analysis is to estimate $\varphi(u_1,u_2 ; x)=\varphi(x)$ where $\varphi(\cdot)$ is a measure of conditional dependence, when a nonparametric consistent estimator of $\varphi(x)$ is not available. We propose to work on a transformation of $\varphi(x)$, encouraging normality, e.g. the Fisher's transformation which maps 
indices defined in $[-1,1]$, like Spearman's $\rho$ or Kendall's $\tau$, into the real line:
\begin{equation}
\label{eq:fishtr}
Z(x) = \frac{1}{2}\log \frac{1+\varphi(x)}{1 - \varphi(x) }.
\end{equation}
 The sample version of $Z(x)$ is then defined by computing the (possibly biased) estimates of $\varphi$:
\begin{equation*}
W(x) = \frac{1}{2} \log \frac{1 + \hat{\varphi}(x)}{1-\hat{\varphi}(x)}.
\end{equation*}

\subsection{Gaussian processes approach}
\label{sub:GP}
First,  we assume that $Z(x)$ follows - \textit{a priori} - a GP
\begin{equation}
Z(x) \sim \mathcal{GP} \left(  \mathbf{g}(x)^T \boldsymbol{\beta}, \sigma^2 \mathcal{K}(x, x^{\prime}; \xi)  \right).
\label{eq:z}
\end{equation}
Here, the location parameter of the Gaussian process is
\begin{equation*}
\mathbb{E}[Z(x)] = \mathbf{g}(x)^T  \boldsymbol{\beta},   
\end{equation*}
where $\mathbf{g}(x) = \left( g_1 (x), \ldots, g_q (x)  \right)^T $ is a set of known functions, $x \in \mathbb{R}^p$ and $\boldsymbol{\beta} \in \mathbb{R}^q$.
Common choices for the basis function $\mathbf{g}(x)$ are $\mathbf{0}$, $(1,\mathbf{x})$ or 
$(1,\mathbf{x},\mathbf{x}^2)$, and so on.

\noindent
Also, $\mathcal{K}(x, x^\prime; \xi)$ is a generic correlation kernel depending on a parameter $\xi$, and $\sigma^2$ is a positive scale parameter, so that
\begin{equation*}
\mathbb{C}ov(Z(x),Z(x^{\prime})) = \sigma^2 \mathcal{K}(x,x^{\prime}; \xi).
\end{equation*}
Without loss of generality, we will consider the squared exponential kernel
\begin{equation*}
\mathbb{C}ov \left( Z(x), Z(x^{\prime}) \right) =
\sigma^2 \mathcal{K}(x, x^{\prime}; \xi)  = \sigma^2 \exp \left( - \frac{1}{2}\sum_{\iota=1}^{m}\frac{d(x_{\iota}, x_{\iota}^{\prime})}{\xi_\iota} \right) 
 = \sigma^2 \exp \left( - \frac{1}{2}\sum_{\iota=1}^{m}\frac{(x_{\iota} - x_{\iota}^{\prime})^2}{\xi_\iota} \right). 
\end{equation*}
For ease of notation, we will consider the case where $m=1$, that is
\begin{equation*}
\mathcal{K}(x, x^{\prime}; \xi) = \exp \left( - \frac{1}{2} \frac{(x - x^{\prime})^2}{\xi} \right);
\end{equation*}
however, generalizations to $m>1$ are straightforward. 

We estimate the functional $\varphi(x_\ell)$ using the unconditional consistent estimator of $\varphi$, say $\hat{\varphi}$ and assume that each set of observations, $(y_{1i},y_{2i},x_\ell)$, $i=1,\ldots,n_\ell$, generates a noisy version of $Z(x_\ell)$, say $W(x_\ell)$, which depends on a statistic evaluated at location $x_\ell$; in practice, $W(x)$ is a noisy observation of the signal $Z(x)$. 
It is possible
to explicitly model the noise through some parametric assumption.
For instance, the case of compensating errors can be modelled through a Gaussian distribution
\begin{equation*}
W(x_\ell) = Z(x_\ell) + \varepsilon_\ell    \qquad   \ell = 1, \ldots, k
\end{equation*}
where $Z(x_\ell)$ is defined as in Equation \eqref{eq:z} and $\varepsilon_\ell \sim \mathcal{N}(0, \tau^2_\ell)$, 
with $\tau^2_\ell = {\tau^2}/{n_\ell}$ and 
$\varepsilon_\ell \independent \varepsilon_{\ell^{\prime}}, \ell \not= \ell\prime$.

Consequently, the observations follow a normal distribution
\begin{equation}
W(x_\ell) \sim \mathcal{N} \left( \mathbf{g}(x_\ell)^T \boldsymbol{\beta}, \sigma^2 + \tau^2_\ell  \right) .
\label{eq:w_GP}
\end{equation}
From Equation \eqref{eq:w_GP}, it follows that the likelihood associated to the observations related to locations $x_1, \ldots, x_k$, with a number of observations $n_1, \ldots, n_k$ each, is
\begin{equation*}
L(\boldsymbol{\beta}, \sigma^2, \xi, \tau^2) = \mathcal{N} \left( \mathbf{g}(\bf{x})^T \boldsymbol{\beta}, \sigma^2 \boldsymbol{\Sigma}_{\xi} + \tau^2 \tilde{\mathbf{I}} \right)
\end{equation*}
where $\tilde{\mathbf{I}} = \diag \left( \frac{1}{n_1}, \ldots, \frac{1}{n_k} \right)$ is a diagonal matrix with element $\left( \frac{1}{n_1}, \ldots, \frac{1}{n_k} \right)$ on the diagonal, and
\begin{equation*}
\boldsymbol{\Sigma}_{\xi} = 
 \begin{pmatrix}
  \mathcal{K}(x_1, x_1;\xi) &  \cdots & \mathcal{K}(x_1, x_k;\xi) \\
  \vdots  &  \ddots & \vdots  \\
  \mathcal{K}(x_k, x_1;\xi) &  \cdots & \mathcal{K}(x_k, x_k;\xi) 
 \end{pmatrix}.
\end{equation*}

Standard Bayesian approaches to Gaussian processes can then be developed to handle this situation \citep{goldstein2007bayes,cressie1992statistics}. 
First, the variance matrix can be reformulated following \cite{paulo2005default}:
\begin{equation*}
\sigma^2 \boldsymbol{\Sigma}_{\xi} + \tau^2 \tilde{\mathbf{I}} = \sigma^2 \left( \boldsymbol{\Sigma}_{\xi} + \frac{\tau^2}{\sigma^2} \tilde{\mathbf{I}} \right) = \sigma^2 \left( \boldsymbol{\Sigma}_{\xi} + \lambda \tilde{\mathbf{I}} \right) = \sigma^2 \mathbf{M},
\end{equation*}
where $\lambda=\tau^2 / \sigma^2$ and $\mathbf{M}$ is a positive definite matrix depending on parameters $\xi$ and $\lambda$.

Then, the likelihood can be rewritten as
\begin{equation}
L(\boldsymbol{\beta}, \sigma^2, \lambda, \xi) = \frac{1}{\sigma^k \mid \mathbf{M} \mid^{1/2}} \exp \left( -\frac{1}{2 \sigma^2} (\mathbf{W}-\mathbf{\tilde{X}} \boldsymbol{\beta})^T \mathbf{M}^{-1} (\mathbf{W}-\mathbf{\tilde{X}}\boldsymbol{\beta}) \right),
\label{eq:new_lik}
\end{equation}
where $\mathbf{\tilde{X}} = \mathbf{g}(\bf{x})^T$ is a $k \times q$ matrix of known constants.

Integrating Equation \eqref{eq:new_lik} with respect to $\boldsymbol{\beta}$ (using a noninformative prior distribution $\pi(\boldsymbol{\beta}) \propto 1$), the integrated likelihood function for the variance parameters is obtained as
\begin{align*}
L^I(\sigma^2, \lambda, \xi) &= \int_{B} L(\boldsymbol{\beta}, \sigma^2, \lambda, \xi) \pi(\boldsymbol{\beta}) d\boldsymbol{\beta} \\ 
&\propto \sigma^{-(k-q)} \mid \mathbf{M} \mid ^{-1/2} \mid \mathbf{\tilde{X}}^T \mathbf{M}^{-1} \mathbf{\tilde{X}} \mid^{-1/2} \exp \left( -\frac{1}{2 \sigma^2} \tilde{\mathbf{S}}_{\xi}^2 \right)
\end{align*}
where $\tilde{\mathbf{S}}_{\xi}^2 = \mathbf{W}^T \tilde{\mathbf{Q}} \mathbf{W}$, $\tilde{\mathbf{Q}} = \mathbf{M}^{-1} \tilde{\mathbf{P}}$, $\tilde{\mathbf{P}} = \mathbf{I} - \mathbf{\tilde{X}}(\mathbf{\tilde{X}}^T \mathbf{M}^{-1} \mathbf{\tilde{X}})^{-1}\mathbf{\tilde{X}}^T \mathbf{M}^{-1}$, and $\mathbf{I}$ is the identity matrix. 

It is also possible to integrate with respect to the scale parameter $\sigma^2$, assuming, \textit{a priori}, that it follows an inverse gamma distribution with shape parameter $\alpha$ and scale parameter $r$
\begin{equation*}
\pi(\sigma^2) \propto \frac{1}{\sigma^{\alpha + 1}} \exp \left( - \frac{r}{2 \sigma^2} \right),
\end{equation*}
such that the integrated likelihood for $\xi$ and $\lambda$ is defined as
\begin{align}
L^I(\xi, \lambda) &= \int_0^{\infty} L^I(\sigma^2, \lambda, \xi)\pi(\sigma^2) d\sigma^2 \nonumber \\
&\propto \mid \mathbf{M} \mid^{-1/2} \mid \mathbf{X}^T \mathbf{M}^{-1} \mathbf{X} \mid^{-1/2} \frac{1}{\left( \tilde{\mathbf{S}}_{\xi}^2 + r \right)^{\frac{n-q}{2}+\alpha}}.
\label{eq:inte_lik_xi}
\end{align}
For proofs of the above definition of the integrated likelihood functions, see \cite{paulo2005default}. Expression \eqref{eq:inte_lik_xi} can be also interpreted as the joint density of the observations $(w_1, \ldots, w_k)$ conditionally on
the hyperparameters $\xi, \lambda$. 

In practice, the GP modelling assumptions may not hold (e.g. the bias of the estimator can be asymmetric, the tails may be non-Gaussian, the variance may vary over the parameter space). In order to consider the case of a varying covariance structure depending on the parameter space, we may define 
\begin{equation*}
\mathbf{W} \sim \mathcal{N}\left(\mathbf{g}(\mathbf{x})\boldsymbol{\beta} , \sigma^2\boldsymbol{\Sigma}_{\xi} + \tau^2 \tilde{\mathbf{I}} \exp(\varsigma(\mathbf{x})) \right)
\end{equation*}
where $\varsigma(\mathbf{x})$ is again modelled as a GP: $\varsigma(\mathbf{x}) \sim \mathcal{GP}(\mathbf{0}, \mathcal{K^{*}}(x,x^{\prime}))$. The variance of the Fisher's transform of the copula functional is allowed to vary smoothly as a function of the covariate; the exponential function is used since the variance needs to be positive, then the logarithm is modelled as a GP. Again, $\mathcal{K^{*}}$ can be chosen to be the squared exponential kernel. The covariance matrix may be reparametrized by fixing either $\sigma^2$ or $\tau^2$ equal to one in order to make the model identifiable. Model selection techniques can be used to identify the model to be used; see, for example, \cite{vehtari2012survey}. 

The posterior sample of $(\xi,\lambda)$ is then used to approximate the posterior predictive distribution of $W$ at new locations $x^*$, using the following expression:
\begin{eqnarray*}
f \left( w^\ast | x^*, \bm{W} \right) & = & \int_{\xi, \lambda} f \left( w^\ast, \xi, \lambda | x^*, \bm{W} \right) d \xi d \lambda \\
& = & \int_{\xi, \lambda} f \left( w^\ast | x^*, \xi, \lambda \right) f(\xi, \lambda | \bm{W}) d \xi d \lambda.
\end{eqnarray*}

This approximation can be computed for each value $x^*$ defined over a grid. A natural summary value for the predictive distribution of $W^*$ is the expected value $\mathbb{E}(W | x^*)\approx \bar{W}^*(x^*) = \sum w^* f(w^* | x^*)$, which is also a point estimate of $W^* = Z^* + \varepsilon^* \approx Z^*$.

Finally, the estimation of the functional of interest $\varphi$ can be obtained as
\begin{equation}
\hat{\varphi}(x^*) = \frac{\exp\{2 \bar{W}^*(x^*)\}-1}{\exp\{2 \bar{W}^*(x^*)\}+1}.
\label{eq:phi_hat}
\end{equation}
Alternatively, a full sample of $\varphi(x^*)$ can be derived from the posterior predictive distribution of $W^*(x^*)$.

\subsection{Empirical likelihood approach}
\label{sub:EL}

The assumption of normality of the Fisher's transform of the copula functionals may be admittedly too strict in many situations, in particular because the bias of the conditional copula function estimator is not analytically known. 

In recent years, there has been an interest in finding ways to derive the posterior distribution of the parameters of a model by substituting the likelihood function with an approximation. 
In this setting, \cite{price2018bayesian} propose to use the approximation provided by a synthetic likelihood, for which the distribution of some (not necessarily sufficient) summary statistics of the model is assumed to be Gaussian. Elsewhere \citep{gutmann2016bayesian} simulator-based models are used, which compare the observed datasets with datasets generated from the model; the likelihood function is then approximated by assuming a specific model -- for example, a Gaussian distribution -- for the discrepancy between observed and simulated data (possibly evaluated in terms of summary statistics). 
Another proposal is available in \cite{mengersen2013bayesian}, where the empirical likelihood is employed as a nonparametric approximation of the likelihood function of the parameter of interest. See \cite{grazian2020review} for a recent review of these approaches. 

Recently, \cite{grazian2017approximate} have introduced the use of the empirical likelihood in the specific setting of copula models, by proposing a semiparametric procedure where the posterior distribution of (low-dimensional) functionals of the dependence is derived and the structure of dependence of the joint multivariate distribution is taken as a nuisance parameter defined on a infinite-dimensional space. This approach has two main advantages: i) in many settings, the interest lies in particular indices of the dependence (Spearman's $\rho$, Kendall's $\tau$ or tail dependence indices) which may be in complex relationship with the parameters of the copula function and therefore it can be difficult to derive a likelihood function for them and ii) the selection of a specific copula family can be difficult in applied contexts and a semiparametric approach would avoid the need of choosing among alternative copula models. 

Suppose the model $F_{\mathbf{Y},\psi}(\cdot)$ as expressed in \eqref{eq:uncond_joint} for a multivariate random variable $\mathbf{Y}$ is indexed by a parameter $\psi$ which can be defined as $\psi = (\varphi , \nu)$ and the interest of the analysis is in $\varphi$ while $\nu$ is considered as a nuisance parameter. Then the goal of the analysis is to derive the posterior distribution of $\varphi$
\begin{equation*}
\pi(\varphi | \mathbf{y}) = \int_{N} \frac{f(\mathbf{y} | \nu, \varphi) \pi(\nu| \varphi) \pi(\varphi)}{m(\mathbf{y})} d\nu
\end{equation*}
where $\varphi \in \Phi$ and $\nu \in N$, $\pi(\nu \vert \varphi)\pi(\varphi)$ is the joint prior distribution of $(\varphi, \nu)$ and $m(\mathbf{y})$ is the distribution of the data $\mathbf{y}$ marginalized over the parameter space. 

In the specific setting of this work, we assume that the copula $C(\cdot)$ is parametrized by  $(\varphi, C^*)$
where $\varphi$ is some functional of the dependence in the joint distribution and $C^*$ belongs to an infinite dimensional metric space $(H,d_H)$. 
The empirical likelihood, in particular its Bayesian exponentially tilted version proposed by \cite{schennach2005bayesian}, $L_{BEL}$, may be seen as the derivation of the integrated likelihood for $\varphi$
\begin{equation*}
L_{BEL}(\varphi;\mathbf{y}) \propto \int_{N} L(\varphi,\nu ;\mathbf{y}) \pi(\nu \vert \varphi) d\nu 
\end{equation*}
where the prior distribution $\pi(\nu \vert \varphi)$ is a stochastic process constructed in such a way to give preference to distributions with a large entropy; 
following \cite{schennach2005bayesian}, the empirical likelihood is defined in terms of a set of weights $\{\omega_i\}_{i=1}^n$ obtained as the solution (for each fixed value of 
$\varphi$), of the maximization problem
\begin{equation*}
\max_{(\omega_1,\dots,\omega_n)} \sum_{i=1}^n \left( -\omega_i \log \omega_i \right)
\end{equation*}
under the constraints 
$0 \leq \omega_i \leq 1$ for $i=1,\ldots,n$, $\sum_{i=1}^n \omega_i = 1$, and a moment constraint 
$$\sum_{i=1}^n q(\mathbf{y}_i, \varphi) \omega_i = 0,$$
 for some function $q$.
The approximate Bayesian procedure for deriving an approximate posterior distribution for $\varphi$ is then defined by i) selecting a prior distribution $\pi(\varphi)$; ii) selecting a nonparametric estimator $\hat{\varphi}$, which must be, at least, asymptotically unbiased; iii) computing the empirical likelihood $L_{BEL}(\varphi)$;  iv) deriving the posterior distribution $\pi(\varphi | \mathbf{y})$ through a simulation process. Unfortunately, it is not always easy to perform step ii), i.e. to find an  
(at least, asymptotically) unbiased estimator of the quantity of interest, in order to satisfy the moment condition. In the setting of conditional functional of the dependence, as we have stated in Section \ref{sec:intro}, nonparametric estimators of the copula function are biased, then a fully nonparametric approach cannot be directly implemented by using the empirical likelihood approximation, along the lines of \cite{grazian2017approximate}. 

In this setting, the Fisher's transform of the observations can be defined as a function of the covariates, through a Taylor's expansion in terms of a polynomial of degree $p$. Assume that $W(\cdot)$ is differentiable $p$ times in a neighbourhood of an interior point $x_0$; then
\begin{equation}
W(x_h) \approx W(x_0) + W(x_0)^{\prime} (x_h - x_0) + \ldots + \frac{W(x_0)^{(l)} }{p!}(x_h - x_0)^p \equiv \mathbf{x}^{*T}_{h,x_0} \boldsymbol{\beta}
\label{eq:w_taylor}
\end{equation}
where $\mathbf{x}^{*T}_{h,x_0} = ( 1, (x_h - x_0) , \ldots , (x_h - x_0)^p)$ and $\boldsymbol{\beta} = (\beta_0 , \beta_1, \ldots, \beta_{p})$.

Similarly, a spline approximation, along the lines of \cite{craiu2012mixed}, can be adopted, using a cubic spline as a model for the calibration function of the copula parameter
\begin{equation}
W(x_h) \approx \sum_{\iota=1}^3 \alpha_\iota x_h^\iota + \sum_{s=1}^S \delta_s 
(x_h - \gamma_s)^3_+,
\label{eq:cspline}
\end{equation}
where $a_+ = \max(0,a)$ and $\{\gamma_s\}_{s=1}^S$ is a set of knots. In the approach of \cite{craiu2012mixed}, the copula family is chosen through a model selection procedure.  
%
Then, consistent estimators of the coefficients of the Taylor's expansion or of the spline function, which can be used to define the moment condition in the definition of the empirical likelihood, can be derived. 

For the parameters $\boldsymbol{\beta}$ of the linearised model, it is common to define weakly informative priors, such as, for instance, $\mathcal{N}(0,100)$, and combine them with the weights defined by the empirical likelihood approach described in \cite{grazian2017approximate} in order to obtain a sample of size $G$ of $(\beta_0,\ldots, \beta_{l} | w_1, \ldots, w_k)$ such that
\begin{equation*}
\begin{matrix}
\beta_0^{(1)},\ldots, \beta_{l}^{(1)}, \omega^{(1)} \\
\vdots \\
\beta_0^{(G)}, \ldots, \beta_{l}^{(G)}, \omega^{(G)}
\end{matrix}
\end{equation*}
where $\omega_g$, with $g = 1, \ldots, G$, represents the weights of $(\beta_0^{(g)},\ldots, \beta_{l}^{(g)})$ induced by the empirical likelihood. 
The combinations of $(\boldsymbol{\beta}^{(1)},\ldots,\boldsymbol{\beta}^{(G)})$ and the weights $(\omega_1,\ldots,\omega_G)$ produces an approximation of the joint posterior distribution for $\boldsymbol{\beta}$. 
On the other hand, if a cubic spline as in Equation \eqref{eq:cspline} is used,
weakly informative priors for the parameters $\{\alpha_\iota\}_{\iota=1}^3$ and $\{\delta_s\}=_{s=1}^S$ can be considered, such as, for example, $\mathcal{N}(0,100)$.

The approximate posterior distribution of $\boldsymbol{\beta}$ can be used to approximate the posterior predictive distribution 
\begin{eqnarray*}
f \left( w^\ast | x^*, \bf{W} \right) & = & \int_{\bm{B}} f \left( w^\ast, \boldsymbol{\beta} | x^*, \bm{W} \right) d \boldsymbol{\beta} \\
& = & \int_{\bm{B}} f \left( w^\ast | x^*,  \boldsymbol{\beta}\right) f(\boldsymbol{\beta} | \bm{W}) 
d \boldsymbol{\beta} \\
& \approx & \sum_{g=1}^G  f \left( w^\ast | x^*,  \boldsymbol{\beta}^{(g)}\right) \bar{\omega}^{(g)}
\end{eqnarray*}
where $\bar{\omega}^{(g)} = \frac{\omega^{(g)}}{\sum_{g=1}^G \omega^{(g)}}$ is the normalized weight for the $g$-th iteration. As in Section \ref{sub:GP}, the approximation can be computed on a grid of possible values of $x^*$ and the estimator of the functional of interest can be derived as in Equation \eqref{eq:phi_hat}. 

\subsection{Bayesian splines approach}
\label{sub:splines}

An alternative to the empirical likelihood approach of Section \ref{sub:EL}, still avoiding parametric assumptions, is to model the functional of the dependence as a regression function using regression splines. In this setting, it is possible to introduce assumptions on the type of function, i.e. its shape and degree of smoothness. 

Again, it is useful to work with a transformation of the functional of the dependence, as defined in Equation \eqref{eq:fishtr}:
\begin{equation*}
Z(x) = f(x) + \varepsilon
\end{equation*}
where $f$ is assumed to be smooth; in addition, constraints to force the function to be monotone or convex can be included. 

For a given set of $m$ knots, $a=d_0 = \ldots = d_{k+1} = b$ with $m=k+2$, spline basis functions $\boldsymbol{\delta}=(s_1(x),\ldots,s_m(x))$ are defined on $[a,b]$. Possible choices of splines are quadratic $I$-splines \citep{ramsay1988monotone}, cubic $I$-splines, and $C$-splines \citep{meyer2008inference}.

The function $f(\cdot)$ of the predictor $x$ is modelled as
\begin{equation*}
f(x) = \sum_{j=1}^m \beta_j \boldsymbol{\delta}_j.
\end{equation*}
The coefficients $\beta_j$ are supposed to follow a priori a normal distribution with zero mean and large variance $M$. In addition, a vague gamma prior distribution can be assumed for the precision of the error term. For a full description of the method used in the implementation, we refer the reader to \cite{meyer2011bayesian}

\section{Simulation Study}
\label{sec:simu}
We have performed a simulation study considering one and two covariates. 

\subsection{One covariate}
\label{sub:1cov}

Values of the covariate were simulated from a suitable uniform distribution and the 
Spearman's $\rho$ was associated to the covariate through a linear or a sine relationship
\begin{itemize}
	\item $\rho = 0.8 x -2$ with $x \sim \mathcal{U}\mbox{nif}(2,5)$
	\item $\rho = \sin(x)$ with $x \sim \mathcal{U}\mbox{nif}(-5,5)$
\end{itemize}

For each model, we had $20$ levels of the covariate. 
We simulated two-dimensional observations from four different copula models: Clayton, Frank, Gumbel and Gaussian. For each level of the covariate, we simulated 100 data points with specific parameter values.

Each simulation set-up was repeated 50 times. We first performed a model selection approach based on the \texttt{xvCopula} function of the \texttt{R} \texttt{copula} package \citep{kojadinovic2010modeling}, performing the leave-one-out cross-validation for a set of hypothesized parametric copula models, using maximum pseudo-likelihood estimation. Specifically, we considered,  as potential generating models,  the following copulas: Clayton, Frank, Gumbel, Joe, Gaussian, Plackett, and Student $t$. Table \ref{tab:copula_selec} and \ref{tab:copula_selec_sin} show the absolute frequencies of correct selection of the copula model, when $\rho$ is defined as a linear function and as a sine function respectively. It is evident that the selection of the copula model is not expected to be an easy task. On the other hand, methods based on an assumption about the copula model describing the data, strongly rely on this assumption for inference on the functionals of dependence, such as Spearman's $\rho$, Kendall's $\tau$, or tail dependence indices. Moreover, when selecting the copula based on the observed data, one often assumes that the copula function does not change with the level of the covariate; conversely, when this possibility is taken into account, the number of observations for each level of the covariate are limited, and the selection problem becomes more difficult. 

\begin{table}[h]
\begin{tabular}{c|c|c|c|c|c|c|c}
& \textit{Clayton} & \textit{Frank} & \textit{Gumbel} & \textit{Joe} &\textit{Gaussian} & \textit{Plackett} & \textit{Student t} \\ \hline
Clayton    &    42    &   0  &   0   &  0   &   0   &   1   &  7   \\
Frank      &     0    &  10  &   0   &  0   &   0   &  38   &  3   \\
Gumbel     &     0    &   0  &  39   &  0   &   0   &   0   & 11   \\
Gaussian   &     0    &   0  &   0   &  0   &   8   &   0   & 42   \\
\end{tabular}
\caption{Absolute frequencies of correct selection of the copula model using leaving-one-out cross-validation, where $\rho$ is a linear function of the covariate. The rows represents the copula the data have been simulated from, and the columns the possible models.}
\label{tab:copula_selec}
\end{table} 

\begin{table}[h]
\begin{tabular}{c|c|c|c|c|c|c|c}
& \textit{Clayton} & \textit{Frank} & \textit{Gumbel} & \textit{Joe} &\textit{Gaussian} & \textit{Plackett} & \textit{Student t} \\ \hline
Clayton    &    39    &   0  &   0   &  0   &   0   &   0   & 11   \\
Frank      &     5    &   1  &   7   & 21   &   1   &   0   & 15   \\
Gumbel     &     0    &   0  &   6   &  0   &   0   &   0   & 44   \\
Gaussian   &     1    &   0  &   6   & 27   &   0   &   0   & 16   \\
\end{tabular}
\caption{Absolute frequencies of correct selection of the copula model using leaving-one-out cross-validation, where $\rho$ is a sine function of the covariate. The rows represents the copula the data have been simulated from, and the columns the possible models.}
\label{tab:copula_selec_sin}
\end{table} 

Tables \ref{tab:sim_study_uni}, \ref{tab:sim_study_uni_tau}, \ref{tab:sim_study_uni_sin} and \ref{tab:sim_study_uni_sin_tau} show the results of the comparison among the four approaches described in Section \ref{sec:method}: GPs, the empirical likelihood method based on an inconsistent estimator of the copula, the empirical likelihood method based on a linearised model for the functional of the dependence, and Bayesian splines. 
An approximation of the integrated mean squared error (IMSE) is used to compare the approaches, where the integration is taken with respect to all the possible samples and all the possible covariate levels:
\begin{equation*}
IMSE = \int_{\mathcal{Y}}\int_{\mathcal{X}} (\hat{\rho}(x,y) - \rho(x,y))^2 dx dy \approx \sum_{i=1}^{50} \sum_{\ell=1}^k (\hat{\rho}(x_\ell,\mathbf{y}_i) - \rho(x_\ell,\mathbf{y}_i))^2
\end{equation*}
where $\rho(x_\ell,\mathbf{y}_i)$ is the true value of Spearman's $\rho$ for covariate level $x_\ell$, evaluated at the sample value $\mathbf{y}_i = (y_{1i}, \ldots, y_{ni})$ and $\hat{\rho}(x_\ell,\mathbf{y}_i)$ is the corresponding estimate obtained with each of the four approaches under analysis. 
An analogous expression for the IMSE can be obtained for Kendall's $\tau$.
In addition to the IMSE values, Tables \ref{tab:sim_study_uni}, \ref{tab:sim_study_uni_tau}, \ref{tab:sim_study_uni_sin} and \ref{tab:sim_study_uni_sin_tau} list the average length and average coverage results for the credible intervals of level 95\%. 

Table \ref{tab:sim_study_uni} displays the results with Spearman's $\rho$ as linear function $\rho(x) = 0.8 x -2$, while Table \ref{tab:sim_study_uni_tau} lists the results with Kendall's $\tau$ as linear function $\tau(x) = 0.8 x -2$.
The approaches showing the best performance are the GP (Section \ref{sub:GP}) and Bayesian splines (Section \ref{sub:splines}) based approaches, where in all cases, the coverage is close to the expected one and the IMSE is lower than 0.01. The results based on empirical likelihood approaches are always worse than those based on GPs and Bayesian splines. When using the inconsistent estimator \eqref{eq:estim_cond_cop}, the results are relatively similar, independently of the particular choice of the kernel (Gaussian or triweight) or weights (local-linear or Nadaraya-Watson): the interval coverage is constantly lower than the expected one (around 70\%) and the IMSE is around 0.3, definitely larger than the previous cases. Finally, when using the empirical likelihood approach based on the Taylor expansion \eqref{eq:w_taylor}, with empirical likelihood weights on the $\beta$ coefficients, the variance of the estimates increases and the credible intervals cover all the parameter space. 

\begin{table}[]
\begin{tabular}{cc|cccc}
                                                         & \textit{}                  & \textbf{Clayton} & \textbf{Frank} & \textbf{Gumbel} & \textbf{Gaussian} \\ \hline
\multicolumn{1}{c|}{\textbf{GP}}                         & \textit{IMSE}              & 0.009            & 0.003          & 0.004           & 0.006             \\
\multicolumn{1}{c|}{}                                    & \textit{(Ave) CI Length}   & 0.300            & 0.305          & 0.270           & 0.293             \\
\multicolumn{1}{c|}{}                                    & \textit{(Ave) CI coverage} & 0.910            & 1.000          & 0.978           & 0.891             \\ \hline
\multicolumn{1}{c|}{\textbf{Incons. EL - LL, triweight}} & \textit{IMSE}              & 0.260            & 0.297          & 0.276           & 0.267             \\
\multicolumn{1}{c|}{}                                    & \textit{(Ave) CI Length}   & 1.326            & 1.298          & 1.309           & 1.328             \\
\multicolumn{1}{c|}{}                                    & \textit{(Ave) CI coverage} & 0.717            & 0.671          & 0.725           & 0.760             \\ \hline
\multicolumn{1}{c|}{\textbf{Incons. EL - NW, triweight}} & \textit{IMSE}              & 0.277            & 0.282          & 0.262           & 0.260             \\
\multicolumn{1}{c|}{}                                    & \textit{(Ave) CI Length}   & 1.318            & 1.316          & 1.304           & 1.313             \\
\multicolumn{1}{c|}{}                                    & \textit{(Ave) CI coverage} & 0.728            & 0.727          & 0.730           & 0.739             \\ \hline
\multicolumn{1}{c|}{\textbf{Incons. EL - LL, Gaussian}}  & \textit{IMSE}              & 0.298            & 0.283          & 0.274           & 0.261             \\
\multicolumn{1}{c|}{}                                    & \textit{(Ave) CI Length}   & 1.299            & 1.314          & 1.299           & 1.317             \\
\multicolumn{1}{c|}{}                                    & \textit{(Ave) CI coverage} & 0.723            & 0.718          & 0.704           & 0.775             \\ \hline
\multicolumn{1}{c|}{\textbf{Incons. EL - NW, Gaussian}}  & \textit{IMSE}              & 0.280            & 0.290          & 0.302           & 0.230             \\
\multicolumn{1}{c|}{}                                    & \textit{(Ave) CI Length}   & 1.321            & 1.295          & 1.306           & 1.328             \\
\multicolumn{1}{c|}{}                                    & \textit{(Ave) CI coverage} & 0.000            & 0.729          & 0.721           & 0.734             \\ \hline
\multicolumn{1}{c|}{\textbf{EL - linearised model}}      & \textit{IMSE}              & 1.097            & 1.039          & 1.065           & 0.926             \\
\multicolumn{1}{c|}{}                                    & \textit{(Ave) CI Length}   & 2.000            & 2.000          & 2.000           & 2.000             \\
\multicolumn{1}{c|}{}                                    & \textit{(Ave) CI coverage} & 1.000            & 1.000          & 1.000           & 1.000             \\ \hline
\multicolumn{1}{c|}{\textbf{Bayes Splines}}              & \textit{IMSE}              & 0.003            & 0.003          & 0.004           & 0.002             \\
\multicolumn{1}{c|}{}                                    & \textit{(Ave) CI Length}   & 0.166            & 0.169          & 0.155           & 0.164             \\
\multicolumn{1}{c|}{}                                    & \textit{(Ave) CI coverage} & 0.975            & 0.985          & 0.880           & 0.970            
\end{tabular}
\caption{Results for the nonparametric analyses with Spearman's $\rho$ as a linear function. ``GP'' denotes the Gaussian process model of Section \ref{sub:GP}; ``Incons. EL'' denotes the empirical likelihood approach using the inconsistent estimator \eqref{eq:estim_cond_cop}, with Local Linear (LL) or Nadaraya-Watson (NW) weights and triweight or Gaussian kernel; ``EL - linearised model'' denotes the empirical likelihood approach of Section \ref{sub:EL} using the approximation \eqref{eq:w_taylor}; ``Bayes Splines'' denotes the approach of Section \ref{sub:splines}.}
\label{tab:sim_study_uni}
\end{table}

\begin{table}[]
\begin{tabular}{cc|cccc}
                                                         & \textit{}                  & \textbf{Clayton} & \textbf{Frank} & \textbf{Gumbel} & \textbf{Gaussian} \\ \hline
\multicolumn{1}{c|}{\textbf{GP}}                         & \textit{IMSE}              & 0.008            & 0.000          & 0.009           & 0.009             \\
\multicolumn{1}{c|}{}                                    & \textit{(Ave) CI Length}   & 0.187            & 0.174          & 0.217           & 0.217             \\
\multicolumn{1}{c|}{}                                    & \textit{(Ave) CI coverage} & 0.623            & 1.000          & 0.771           & 0.771             \\ \hline
\multicolumn{1}{c|}{\textbf{Incons. EL - LL, triweight}} & \textit{IMSE}              & 0.264            & 0.255          & 0.242           & 0.242             \\
\multicolumn{1}{c|}{}                                    & \textit{(Ave) CI Length}   & 1.269            & 1.303          & 1.337           & 1.337             \\
\multicolumn{1}{c|}{}                                    & \textit{(Ave) CI coverage} & 0.703            & 0.737          & 0.773           & 0.773             \\ \hline
\multicolumn{1}{c|}{\textbf{Incons. EL - NW, triweight}} & \textit{IMSE}              & 0.322            & 0.257          & 0.244           & 0.244             \\
\multicolumn{1}{c|}{}                                    & \textit{(Ave) CI Length}   & 1.340            & 1.320          & 1.319           & 1.319             \\
\multicolumn{1}{c|}{}                                    & \textit{(Ave) CI coverage} & 0.698            & 0.715          & 0.751           & 0.751             \\ \hline
\multicolumn{1}{c|}{\textbf{Incons. EL - LL, Gaussian}}  & \textit{IMSE}              & 0.297            & 0.283          & 0.240           & 0.240             \\
\multicolumn{1}{c|}{}                                    & \textit{(Ave) CI Length}   & 1.291            & 1.324          & 1.311           & 1.311             \\
\multicolumn{1}{c|}{}                                    & \textit{(Ave) CI coverage} & 0.711            & 0.753          & 0.743           & 0.743             \\ \hline
\multicolumn{1}{c|}{\textbf{Incons. EL - NW, Gaussian}}  & \textit{IMSE}              & 0.326            & 0.294          & 0.301           & 0.301             \\
\multicolumn{1}{c|}{}                                    & \textit{(Ave) CI Length}   & 1.369            & 1.278          & 1.243           & 1.243             \\
\multicolumn{1}{c|}{}                                    & \textit{(Ave) CI coverage} & 0.000            & 0.000          & 0.000           & 0.000             \\ \hline
\multicolumn{1}{c|}{\textbf{EL - linearised model}}      & \textit{IMSE}              & 0.986            & 0.735          & 0.966           & 0.966             \\
\multicolumn{1}{c|}{}                                    & \textit{(Ave) CI Length}   & 2.000            & 2.000          & 2.000           & 2.000             \\
\multicolumn{1}{c|}{}                                    & \textit{(Ave) CI coverage} & 1.000            & 1.000          & 1.000           & 1.000             \\ \hline
\multicolumn{1}{c|}{\textbf{Bayes Splines}}              & \textit{IMSE}              & 0.000            & 0.001          & 0.003           & 0.003             \\
\multicolumn{1}{c|}{}                                    & \textit{(Ave) CI Length}   & 0.141            & 0.137          & 0.153           & 0.153             \\
\multicolumn{1}{c|}{}                                    & \textit{(Ave) CI coverage} & 1.000            & 0.967          & 0.917           & 0.917            
\end{tabular}
\caption{Results for the nonparametric analyses with Kendall's $\tau$ as a linear function. ``GP'' denotes the Gaussian process model of Section \ref{sub:GP}; ``Incons. EL'' denotes the empirical likelihood approach using the inconsistent estimator \eqref{eq:estim_cond_cop}, with Local Linear (LL) or Nadaraya-Watson (NW) weights and triweight or Gaussian kernel; ``EL - linearised model'' denotes the empirical likelihood approach of Section \ref{sub:EL} using the approximation \eqref{eq:w_taylor}; ``Bayes Splines'' denotes the approach of Section \ref{sub:splines}.}
\label{tab:sim_study_uni_tau}
\end{table}

Similarly to Tables \ref{tab:sim_study_uni} and \ref{tab:sim_study_uni_tau}, Tables \ref{tab:sim_study_uni_sin} and \ref{tab:sim_study_uni_sin_tau} report the performance study of the different methods when the true functions are $\rho(x)=\sin(x)$ (Table \ref{tab:sim_study_uni_sin}) and $\tau(x)=\sin(x)$ (Table \ref{tab:sim_study_uni_sin_tau}) respectively. Again, the best performance is achieved by the GPs and the Bayesian splines approaches. The Bayesian splines show a coverage which is similar or higher than the one obtained with the GPs approach, with only slightly longer intervals on average. The methods based on empirical likelihood approximations seem inconsistent, with larger intervals lengths and reduced coverage. 

\begin{table}[]
\begin{tabular}{cc|cccc}
                                                    & \textit{}                  & \textbf{Clayton} & \textbf{Frank} & \textbf{Gumbel} & \textbf{Gaussian} \\ \hline
\multicolumn{1}{c|}{\textbf{GP}}                    & \textit{IMSE}              & 0.504            & 0.492          & 0.428           & 0.366             \\
\multicolumn{1}{c|}{}                               & \textit{(Ave) IC Length}   & 1.214            & 1.001          & 1.007           & 1.258             \\
\multicolumn{1}{c|}{}                               & \textit{(Ave) IC coverage} & 0.715            & 0.862          & 0.892           & 0.787             \\ \hline
\multicolumn{1}{c|}{\textbf{Incons. EL - LL, triweight}} & \textit{IMSE}              & 1.620            & 1.972          & 1.762           & 1.672             \\
\multicolumn{1}{c|}{}                                    & \textit{(Ave) CI Length}   & 1.326            & 1.298          & 1.309           & 1.328             \\
\multicolumn{1}{c|}{}                                    & \textit{(Ave) CI coverage} & 0.175            & 0.185          & 0.253           & 0.600             \\ \hline
\multicolumn{1}{c|}{\textbf{Incons. EL - NW, triweight}} & \textit{IMSE}              & 1.217            & 1.823          & 1.612           & 1.601             \\
\multicolumn{1}{c|}{}                                    & \textit{(Ave) CI Length}   & 1.518            & 1.161          & 1.324           & 1.413             \\
\multicolumn{1}{c|}{}                                    & \textit{(Ave) CI coverage} & 0.482            & 0.437          & 0.478           & 0.416             \\ \hline
\multicolumn{1}{c|}{\textbf{Incons. EL - LL, Gaussian}}  & \textit{IMSE}              & 1.281            & 1.238          & 1.245           & 1.621             \\
\multicolumn{1}{c|}{}                                    & \textit{(Ave) CI Length}   & 1.229            & 1.314          & 1.229           & 1.316             \\
\multicolumn{1}{c|}{}                                    & \textit{(Ave) CI coverage} & 0.623            & 0.628          & 0.654           & 0.685             \\ \hline
\multicolumn{1}{c|}{\textbf{Incons. EL - NW, Gaussian}}  & \textit{IMSE}              & 1.202            & 1.270          & 1.202           & 1.236             \\
\multicolumn{1}{c|}{}                                    & \textit{(Ave) CI Length}   & 1.222            & 1.198          & 1.273           & 1.210             \\
\multicolumn{1}{c|}{}                                    & \textit{(Ave) CI coverage} & 0.410            & 0.387          & 0.321           & 0.344             \\ \hline
\multicolumn{1}{c|}{\textbf{EL - linearised model}} & \textit{IMSE}              & 1.437            & 1.071          & 1.242           & 1.417             \\
\multicolumn{1}{c|}{}                               & \textit{(Ave) IC Length}   & 2.000            & 2.000          & 2.000           & 2.000             \\
\multicolumn{1}{c|}{}                               & \textit{(Ave) IC coverage} & 1.000            & 1.000          & 1.000           & 1.000             \\ \hline
\multicolumn{1}{c|}{\textbf{Bayes Splines}}         & \textit{IMSE}              & 0.541            & 0.502          & 0.720           & 0.541             \\
\multicolumn{1}{c|}{}                               & \textit{(Ave) IC Length}   & 1.229            & 1.071          & 1.051           & 1.206             \\
\multicolumn{1}{c|}{}                               & \textit{(Ave) IC coverage} & 0.931            & 1.000          & 0.825           & 0.912            
\end{tabular}
\caption{Results for the nonparametric analyses with Spearman's $\rho$ as a sine function. ``GP'' denotes the Gaussian process model of Section \ref{sub:GP}; ``Incons. EL'' denotes the empirical likelihood approach using the inconsistent estimator \eqref{eq:estim_cond_cop}, with Local Linear (LL) or Nadaraya-Watson (NW) weights and triweight or Gaussian kernel; ``EL - linearised model'' denotes the empirical likelihood approach of Section \ref{sub:EL} using the approximation \eqref{eq:w_taylor}; ``Bayes Splines'' denotes the approach of Section \ref{sub:splines}.}
\label{tab:sim_study_uni_sin}
\end{table}

\begin{table}[]
\begin{tabular}{cc|cccc}
                                                         & \textit{}                  & \textbf{Clayton} & \textbf{Frank} & \textbf{Gumbel} & \textbf{Gaussian} \\ \hline
\multicolumn{1}{c|}{\textbf{GP}}                         & \textit{IMSE}              & 0.645            & 0.293          & 0.563           & 0.494             \\
\multicolumn{1}{c|}{}                                    & \textit{(Ave) CI Length}   & 0.391            & 0.732          & 0.949           & 1.577             \\
\multicolumn{1}{c|}{}                                    & \textit{(Ave) CI coverage} & 0.518            & 0.549          & 0.398           & 0.792             \\ \hline
\multicolumn{1}{c|}{\textbf{Incons. EL - LL, triweight}} & \textit{IMSE}              & 0.144            & 0.135          & 0.350           & 0.131             \\
\multicolumn{1}{c|}{}                                    & \textit{(Ave) CI Length}   & 0.809            & 0.831          & 0.747           & 0.849             \\
\multicolumn{1}{c|}{}                                    & \textit{(Ave) CI coverage} & 0.320            & 0.380          & 0.260           & 0.420             \\ \hline
\multicolumn{1}{c|}{\textbf{Incons. EL - NW, triweight}} & \textit{IMSE}              & 0.805            & 0.809          & 0.797           & 0.882             \\
\multicolumn{1}{c|}{}                                    & \textit{(Ave) CI Length}   & 1.134            & 1.320          & 1.141           & 1.102             \\
\multicolumn{1}{c|}{}                                    & \textit{(Ave) CI coverage} & 0.320            & 0.340          & 0.400           & 0.340             \\ \hline
\multicolumn{1}{c|}{\textbf{Incons. EL - LL, Gaussian}}  & \textit{IMSE}              & 0.303            & 0.305          & 0.457           & 0.309             \\
\multicolumn{1}{c|}{}                                    & \textit{(Ave) CI Length}   & 0.598            & 0.754          & 0.687           & 0.662             \\
\multicolumn{1}{c|}{}                                    & \textit{(Ave) CI coverage} & 0.320            & 0.260          & 0.280           & 0.200             \\ \hline
\multicolumn{1}{c|}{\textbf{Incons. EL - NW, Gaussian}}  & \textit{IMSE}              & 0.669            & 0.597          & 0.806           & 0.605             \\
\multicolumn{1}{c|}{}                                    & \textit{(Ave) CI Length}   & 0.673            & 0.544          & 0.767           & 0.579             \\
\multicolumn{1}{c|}{}                                    & \textit{(Ave) CI coverage} & 0.340            & 0.320          & 0.360           & 0.320             \\ \hline
\multicolumn{1}{c|}{\textbf{EL - linearised model}}      & \textit{IMSE}              & 1.420            & 1.504          & 1.854           & 1.534             \\
\multicolumn{1}{c|}{}                                    & \textit{(Ave) CI Length}   & 2.000            & 2.000          & 2.000           & 2.000             \\
\multicolumn{1}{c|}{}                                    & \textit{(Ave) CI coverage} & 1.000            & 1.000          & 1.000           & 1.000             \\ \hline
\multicolumn{1}{c|}{\textbf{Bayes Splines}}              & \textit{IMSE}              & 0.492            & 0.491          & 0.944           & 0.507             \\
\multicolumn{1}{c|}{}                                    & \textit{(Ave) CI Length}   & 1.039            & 1.003          & 0.517           & 1.066 \\
\multicolumn{1}{c|}{}                                    & \textit{(Ave) CI coverage} & 0.400            & 0.330          & 0.100           & 0.350            
\end{tabular}
\caption{Results for the nonparametric analyses with Kendall's $\tau$ as a sine function. ``GP'' denotes the Gaussian process model of Section \ref{sub:GP}; ``Incons. EL'' denotes the empirical likelihood approach using the inconsistent estimator \eqref{eq:estim_cond_cop}, with Local Linear (LL) or Nadaraya-Watson (NW) weights and triweight or Gaussian kernel; ``EL - linearised model'' denotes the empirical likelihood approach of Section \ref{sub:EL} using the approximation \eqref{eq:w_taylor}; ``Bayes Splines'' denotes the approach of Section \ref{sub:splines}.}
\label{tab:sim_study_uni_sin_tau}
\end{table}


\subsection{Two covariates}

We have performed simulations also considering functionals depending on two covariates, $x_1$ and $x_2$. We remind that some of the methods investigated in this paper require repetitions for each combination of the covariate levels, therefore they are not applicable when using continuous covariates, unless values are grouped into classes. Here, we compare the results with the semi-parametric approach proposed by \cite{levi2018bayesian}: while such method requires the selection of the copula function, it does not require repetitions for each level of the covariates. Here, following \cite{levi2018bayesian}, we use the conditional cross-validated pseudo marginal likelihood (CCVML) as selection tool; the CCVML considers the predictive distribution of one response given the rest of the data. 

We generated again 50 repetitions of simulations for each of the four considered copulas: Clayton, Frank, Gumbel, and Gaussian. The methods can be performed for any functional of the dependence, however we focus here on the Kendall's $\tau$ without loss of generality. 
We fixed
\begin{equation}
\tau = 0.7 + 0.15 \sin (\sqrt{10}(x_1 + 3x_2))
\label{eq:tau_2cov}
\end{equation}
with $x_1, x_2$ independently and identically distributed from $\mathcal{U}\mbox{nif}(0,1)$. 
We adopted the application setting of the algorithm used in Scenario 1 of the paper by \cite{levi2018bayesian}.
We considered 10 repetitions for each combination of the levels of $x_1$ and $x_2$. Table \ref{tab:copula_selec_2} shows the results of the selection task for the copula model, using the CCVML. Again, the identification of the correct copula model is not easy, in particular it seems that the Gaussian copula is poorly identified through CCVML. 

\begin{table}[h]
\begin{tabular}{c|c|c|c|c}
& \textit{Clayton} & \textit{Frank} & \textit{Gaussian} & \textit{Gumbel}  \\ \hline
Clayton    &    26    &   11  &  13   &  0    \\
Frank      &    13    &   15  &  22   &  0    \\
Gumbel     &     0    &    0  &  10   & 40    \\
Gaussian   &     3    &   10  &   1   & 36    \\
\end{tabular}
\caption{Absolute frequencies of correct selection of the copula model using CCVML, where $\tau$ is a function of two covariates, see Equation \ref{eq:tau_2cov}. The rows represents the copula the data have been simulated from, and the columns the possible models.}
\label{tab:copula_selec_2}
\end{table} 

Table \ref{tab:sim_study_two_cov} compares the results of the method proposed by \cite{levi2018bayesian} with the methods presented in Section \ref{sec:method}:
GPs, the empirical likelihood method based on an inconsistent estimator of the copula, the empirical likelihood
method based on a linearised model for the functional of the dependence and Bayesian splines.
The method based on GPs shows the best performance in terms of average IMSE. In this case, Bayesian splines show slightly larger IMSE results than GPs, which may be due to the increased dimensionality of the problem. 

To implement the method based on the inconsistent copula estimator, it is necessary to extend the definition of the Nadaraya-Watson weigths given in Section \ref{sec:method}. Specifically, for $p$ covariates the weights are defined as
\begin{equation}
w_i(\mathbf{x},\mathbf{h}_n) = \frac{\mathcal{K}\left( \frac{\mathbf{X}_i - \mathbf{x}}{\mathbf{h}_n}\right)}{\sum_{i=1}^n \mathcal{K}\left( \frac{\mathbf{X}_i - \mathbf{x}}{\mathbf{h}_n}\right)}
\end{equation}
where $\mathcal{K}(\mathbf{X}_i - \mathbf{x}) = \mathcal{K}_{h_1}(X_{i1} - x_1) \times \ldots \times \mathcal{K}_{h_p}(X_{ip} - x_p)$ for some choices of bandwidth $(h_1, \ldots, h_p)$. Again, we choose a triweight and a Gaussian kernel function. 

Table \ref{tab:sim_study_two_cov} shows that the performance of the empirical likelihood method results in lower IMSE values than those yielded by the conditional method of \cite{levi2018bayesian}, but in larger IMSE values than those obtained with the methods based on GPs and splines. In addition, the implementation of the empirical likelihood approach is computationally intensive, and it requires a limited number of covariates. Finally, as for Section \ref{sub:1cov}, the performance of the method based on the empirical likelihood computed on the linearised model is the worst amongst the analysed methods. 

Here we remind that the approaches based on GPs and Bayesian splines require replications for each combination of covariate levels, while the semi-parametric approach of \cite{levi2018bayesian} and the approaches based on the empirical likelihood can be applied when no replications are available. However, the semi-parametric conditional approach relies on the selection of the copula family (see Table \ref{tab:copula_selec_2}).

\begin{table}[]
\begin{tabular}{c|cccc}
                                   & \textbf{Clayton} & \textbf{Frank} & \textbf{Gumbel} & \textbf{Gaussian} \\ \hline
\textbf{Semi-parametric (LC2018)}  & 0.451            & 0.467          & 0.452           & 0.180             \\
\textbf{GP}                        & 0.002            & 0.002          & 0.003           & 0.004             \\
\textbf{Incons. EL - NW triweight} & 0.041            & 0.041          & 0.041           & 0.041             \\
\textbf{Incons. EL - NW Gaussian} & 0.042            & 0.042          & 0.041           & 0.041             \\
\textbf{EL - linearised model}     & 1.057            & 1.061          & 1.061           & 1.001             \\
\textbf{Bayes Splines}             & 0.009            & 0.009          & 0.009           & 0.009            
\end{tabular}
\caption{Integrated Mean Squared Errors (IMSE) for the nonparametric analyses of the Kendall's $\tau$ in presence of two covariates. ``Semi-parametric (LC2018)'' stands for the method proposed by \cite{levi2018bayesian}, while the other acronyms can be interpreted as in Table \ref{tab:sim_study_uni}.}       
\label{tab:sim_study_two_cov}
\end{table}

\section{Real Data Examples}
\label{sec:app}

We now apply the methods described in this work to two real data examples, the first in the area of civil engineering and the second in astrophysics, to compare the results of the analysis performed by the different methodologies on realistic problems. 

\subsection{Energy Efficiency}
\label{sub:energy}
The Energy Efficiency dataset includes simulations for 12 different building architectural interior environments, obtained with the Ecotect software
 \citep{roberts2001ecotect}. The buildings differ with respect to 8 features which, combined in different ways, lead to 768 building shapes. The response variables are the heating and the cooling loads, i.e. the amount of heat energy that would need to be added or removed to the space to maintain the temperature in the requested range, respectively. Lower thermal loads are indicators of higher energy efficiency. The data have been generated and analysed in \cite{tsanas2012accurate} \footnote{The data are available at \url{https://archive.ics.uci.edu/ml/datasets/Energy+efficiency} }.

\begin{figure}
\centering
\begin{subfigure}{.5\textwidth}
  \centering
  \includegraphics[width=7cm,height=5cm]{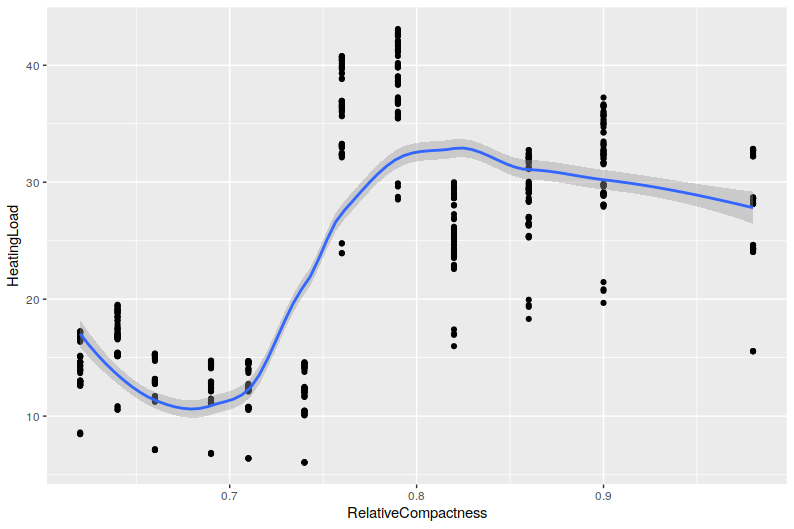}
  \caption{Heating load against relative compactness}
  \label{fig:sub1}
\end{subfigure}%
\begin{subfigure}{.5\textwidth}
  \centering
  \includegraphics[width=7cm,height=5cm]{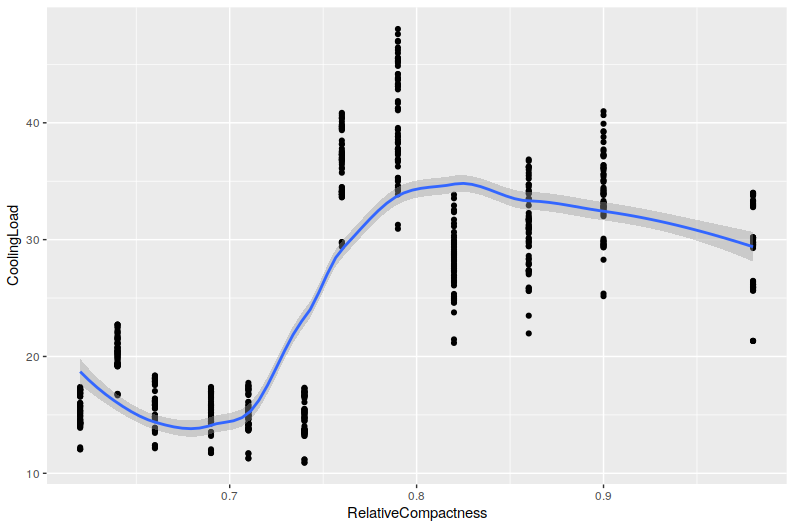}
  \caption{Cooling load against relative compactness}
  \label{fig:sub2}
\end{subfigure}
\caption{Scatterplots of heating and cooling load measures (plotted on the vertical axes) against relative compactness (on the horizontal axes), from the Energy Efficiency dataset. The blue lines are obtained as smoothed conditional means, while the grey areas show the relevant confidence intervals.}
\label{fig:energy_data}
\end{figure}

An important feature when studying the energy efficiency of buildings is the relative compactness (RC), which is a measure to compare different building shapes through the surface to volume ratio \citep{pessenlehner2003building,ourghi2007simplified}. Figure \ref{fig:energy_data} shows the scatterplots of the observed data of heating and cooling loads against RC. The blue lines indicates the smoothed conditional mean, showing that the relationship is non-linear. We now analyse the effect on the dependence between heating and cooling loads with respect to the RC indicator. 

We applied the approaches described in Section \ref{sec:method} to the Energy Efficiency dataset, after estimating each marginal nonparametrically. 

\begin{figure}
  \centering
  \includegraphics[width=12cm,height=7cm]{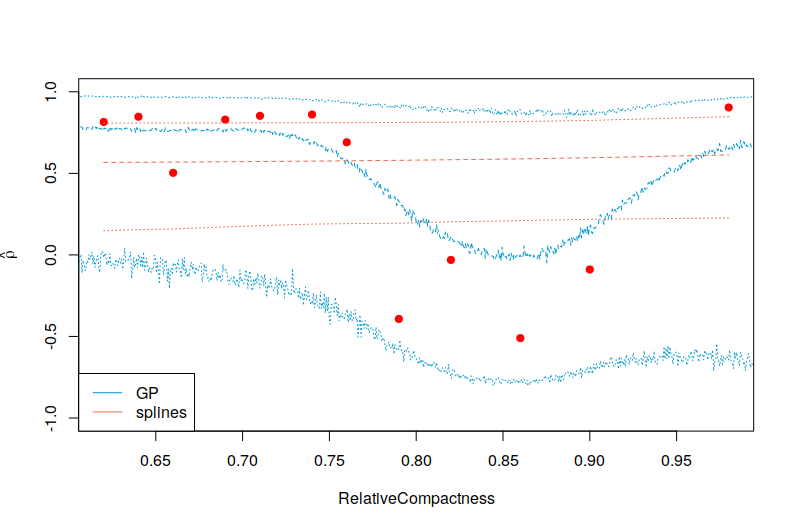}
  \caption{Approximations of Spearman's $\rho$ computed between heating and cooling loads, varying as a function of RC from the Energy Efficiency dataset, obtained with the GP method illustrated in Section \ref{sub:GP} (blue lines) and the splines method illustrated in Section \ref{sub:splines} (coral lines), respectively. 
The inner dashed lines denote the posterior means and the dotted lines denote the 95\% credible intervals. 
The red points are frequentist estimates for each of the levels of RC.}
  \label{fig:energy_GP_splines}
\end{figure}

Figure \ref{fig:energy_GP_splines} shows the approximation of the Spearman's $\rho$ computed between heating and cooling loads as a function of RC. 
The blue lines show the approximation obtained with the GP method illustrated in Section \ref{sub:GP} and the coral lines show results of the splines method illustrated in Section \ref{sub:splines}. 
The inner dashed lines denote the posterior means and the dotted lines denote the 95\% credible intervals. 
The red points are frequentist estimates for each of the levels of RC.
From Figure \ref{fig:energy_GP_splines} it seems that the dependence between heating and cooling loads is strong for low and high levels of RC, respectively. For moderate levels of RC, instead, it seems that the strength of dependence is lower. This may be due to an increased variability in the data, as shown in Figure \ref{fig:energy_data}. A comparison between the method based on GPs and the method based on splines shows that the approximation obtained through the GP method seems to better follow the frequentist estimates of Spearman's $\rho$ (red points in Figure \ref{fig:energy_GP_splines}), while the approximation obtained through splines is less sensitive to changes in the value of the dependence. This may be due to the limited number of points for each level of the covariate, showing that GPs is able to follow changes in dependence with a lower number of data points. 

\begin{figure}
  \centering
  \includegraphics[width=14cm,height=10cm]{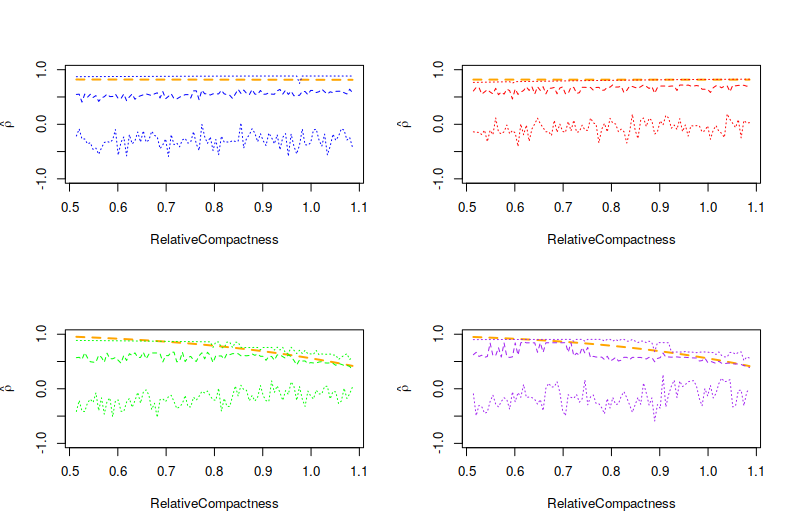}
  \caption{Approximations of Spearman's $\rho$ computed between heating and cooling loads, varying as a function of RC from the Energy Efficiency dataset, obtained with the method based on EL. The orange long-dashed lines denote the frequentist estimates obtained with the inconsistent estimator of the copula, the coloured dashed lines denote the posterior means obtained with the EL methods and the dotted lines denote the 95\% credible intervals. Each figure represents the approximation obtained with NW weights and triweight kernel (top left), NW weights and Gaussian kernel (top right), LL weights and triweight kernel (bottom left), and LL weights and Gaussian kernel (bottom right).}
  \label{fig:energy_EL}
\end{figure}

Figure \ref{fig:energy_EL} shows a similar approximation of the posterior distribution of Spearman's $\rho$ obtained with the method based on the empirical likelihood (EL) illustrated in Section \ref{sub:EL}, using the inconsistent estimator of the copula function. 
The orange long-dashed lines denote the frequentist estimates obtained with the inconsistent estimator of the copula, the coloured dashed lines denote the posterior means obtained with the EL methods and the dotted lines denote the 95\% credible intervals. Each figure represents the approximation obtained with NW weights and triweight kernel (top left), NW weights and Gaussian kernel (top right), LL weights and triweight kernel (bottom left), and LL weights and Gaussian kernel (bottom right).
From Figure \ref{fig:energy_EL} it is evident that the approximation depends on the weights and kernel chosen in the estimator of the copula function. All the approximations tend to be concentrated around large values of dependence, however methods based on local-linear weights show a decline of the dependence for large values of RC. Methods based on the EL for the parameters of the linearised model show highly variable posterior approximations of the parameters, that lead to estimates of Spearman's $\rho$ which are very uncertain (with credible intervals including all the parameter space) and are not included here. 

\begin{figure}
  \centering
  \includegraphics[width=14cm,height=10cm]{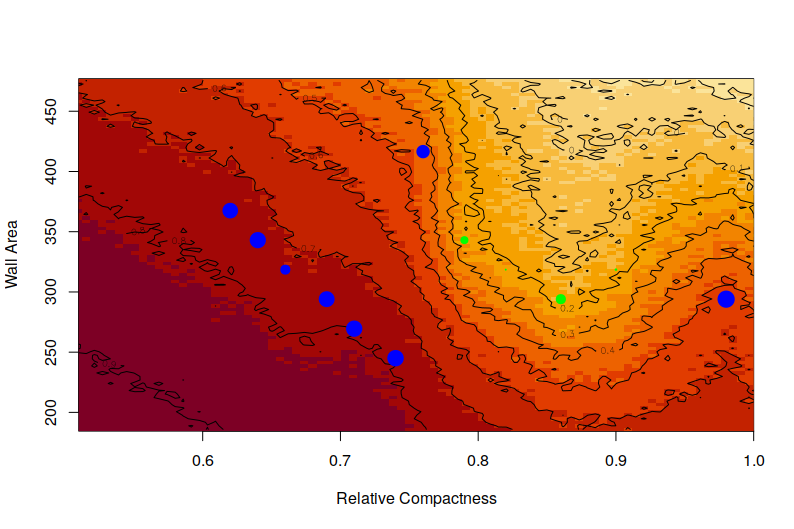}
  \caption{Estimation of Sperman's $\rho$, with respect to the levels of WA (vertical axis) and RC (horizontal axis), for the Energy Efficiency dataset. 
Different colours denote the Sperman's $\rho$ posterior conditional mean levels obtained by applying the GP method.
Dark red denotes strong dependence, while light yellow denotes weak dependence levels. 
  The dots represent the frequentist estimates of the Spearman's $\rho$ computed as unconditional samples for the observations with that particular combination of the covariate levels. Blue dots denote positive $\rho$s and green dots denote negative $\rho$s. The size of the dots represent the scaled absolute value of the estimates.}
  \label{fig:energy_2cov}
\end{figure}

The presence of several building features in the Energy Efficiency dataset allows for the application of methods based on more than one covariate. 
In particular, we now focus the analysis on two covariates: RC and wall area (WA). Table \ref{tab:energy_2cov} shows the absolute frequency of the data for each combination of the RC and WA covariate levels. 
Figure \ref{fig:energy_2cov} shows the Sperman's $\rho$ posterior conditional means obtained by applying the method based on GP, with respect to the levels of WA and RC. 
Dark red denotes strong dependence, while light yellow denotes weak dependence levels.
  The dots represent the frequentist estimates of the Spearman's $\rho$ computed as unconditional samples for the observations with that particular combination of the covariate levels. Blue dots denote positive $\rho$s and green dots denote negative $\rho$s. The size of the dots represent the scaled absolute value of the estimates.
  
Due to the data setting, as illustrated in Table \ref{tab:energy_2cov}, the implementation of the Bayesian splines is made more difficult for the limited amount of data points for each of the few observed combinations of covariates and the method does not converge. 
Similarly, the EL computed on the multivariate versions of the NW or LL weights tends to perform poorly, with many weights being close or equal to zero, which influences the overall approximation of the likelihood function. 

In conclusion, in this example the only method applicable with two covariates is the one based on GPs amongst the methods presented in this work. However, it is clear that, as the number of covariates (or the number of levels in each covariate) increases, the method would require an increasing number of  observations, and this reduces the applicability in high-dimensional settings. 

\begin{table}[]
\begin{tabular}{cc|ccccccc}
& & \multicolumn{7}{c}{\textbf{Wall Area}} \\
      &        & \textit{245.0} & \textit{269.5} & \textit{294.0} & \textit{318.5} & \textit{343.0} & \textit{367.5} & \textit{416.5} \\ \hline
& \textit{0.62} &                &                &                &                &                & 64             &                \\
& \textit{0.64} &                &                &                &                & 64             &                &                \\
& \textit{0.66} &                &                &                & 64             &                &                &                \\
& \textit{0.69} &                &                & 64             &                &                &                &                \\
& \textit{0.71} &                & 64             &                &                &                &                &                \\
\textbf{Relative} & \textit{0.74} & 64             &                &                &                &                &                &                \\
\textbf{Compactness} & \textit{0.76} &                &                &                &                &                &                & 64             \\
& \textit{0.79} &                &                &                &                & 64             &                &                \\
& \textit{0.82} &                &                &                & 64             &                &                &                \\
& \textit{0.86} &                &                & 64             &                &                &                &                \\
& \textit{0.90} &                &                &                & 64             &                &                &                \\
& \textit{0.98} &                &                & 64             &                &                &                &               
\end{tabular}
\caption{Frequency table of Relative Compactness (on the rows) and Wall Area (on the columns) from the Energy Efficiency dataset.}
\label{tab:energy_2cov}
\end{table}

\subsection{MAGIC Gamma Telescope}

The Cherenkov gamma telescope observes high energy gamma rays, detecting the radiation emitted by charged particles produced inside electromagnetic showers. Photons are collected in patterns forming the shower image and it is necessary to discriminate between the image caused by primary gamma rays and the one caused by other cosmic rays. 
Images are usually ellipses and their features, in terms of measures associated with the long and short axes, help to discriminate amongst images. 

\begin{figure}
  \centering
  \includegraphics[height=15cm]{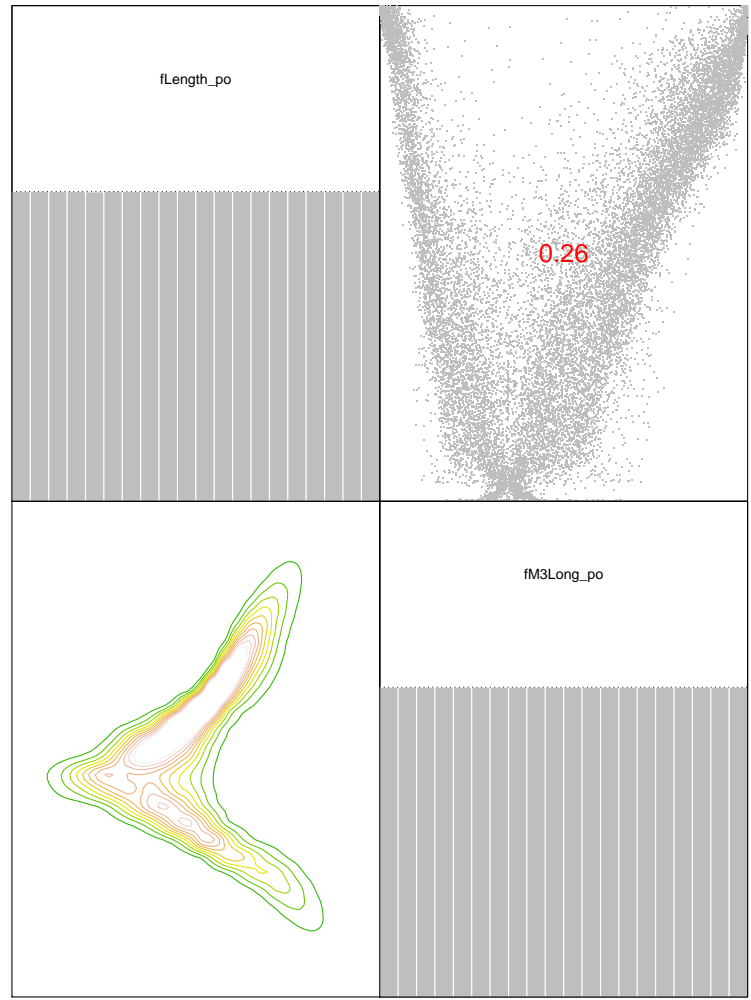}
  \caption{Empirical normalized contour plot of the variables \texttt{Length} and \texttt{M3Long} (bottom left panel) from the MAGIC Gamma Telescope dataset; scatterplot and correlation between the variables (top right panel);
histograms of \texttt{Length} (top left panel) and \texttt{M3Long} (bottom right panel), after transforming the variables into pseudo-observations.}
  \label{fig:contour_magic}
\end{figure}

The data used in this Section are simulations of ellipses parameters generated by the Monte Carlo program Corsika \citep{heck1998corsika} to simulate registrations of high energy gamma particles in a Cherenkov gamma telescope, called MAGIC (Major Atmospheric Gamma Imaging Cherenkov) telescope located on the Canary islands. For a full description of the dataset and the evaluation of the performance of several classification methods applied to the data, the reader is referred to \cite{bock2004methods} and \cite{dvovrak2007softening} \footnote{The data are available at \url{https://archive.ics.uci.edu/ml/datasets/magic+gamma+telescope} }.
The dependence between the MAGIC Gamma Telescope variables was analysed by \cite{czado2019analyzing} and by \cite{nagler2016evading}, who pointed out the uncommon characteristics of the dependence structure between some of the variables, which do not correspond to any parametric copula families.
In particular, the dependence between the variables \texttt{Length} (length of the major axis of the ellipse, in mm) and \texttt{M3Long} (third root of the third moment along the major axis, in mm) is rather peculiar, as confirmed by the empirical normalized contour plot depicted in the bottom left panel of Figure \ref{fig:contour_magic}.

Here we analyse how the dependence between \texttt{Length} and \texttt{M3Long}, measured by Spearman's $\rho$, varies with respect to the variables: \texttt{class} (which has two levels: gamma rays or background noise), \texttt{Width} (the length of the minor axis of the ellipse in mm) and \texttt{Size} (the 10-log of the sum of the content of all pixels in the image).

\begin{figure}
\centering
\begin{subfigure}{.5\textwidth}
  \centering
  \includegraphics[width=7cm,height=5cm]{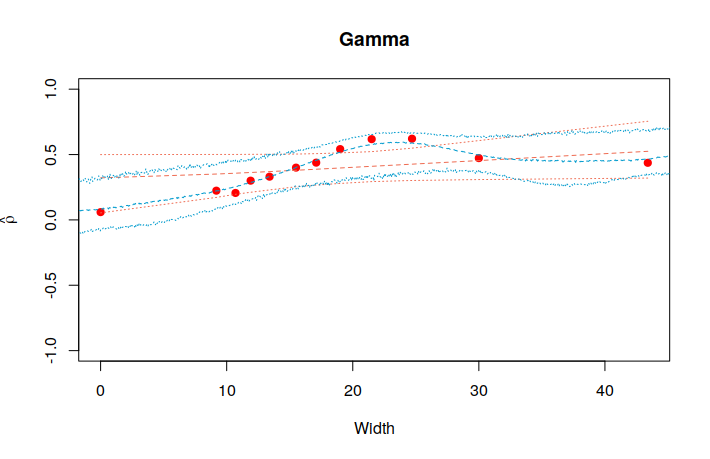}
  \caption{}
\end{subfigure}%
\begin{subfigure}{.5\textwidth}
  \centering
  \includegraphics[width=7cm,height=5cm]{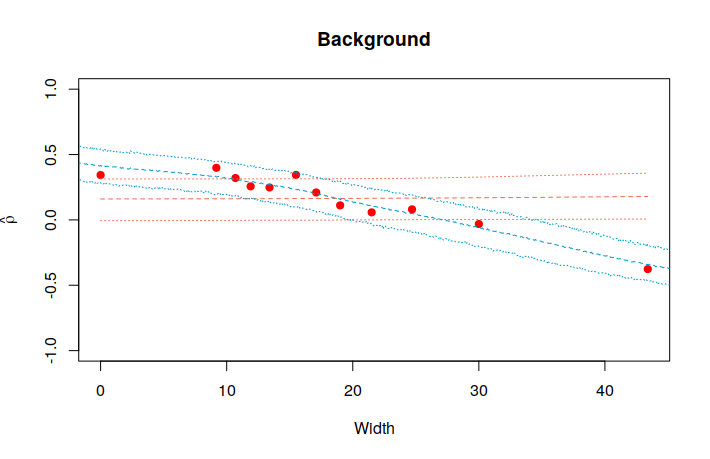}
  \caption{}
\end{subfigure}
\begin{subfigure}{.5\textwidth}
  \centering
  \includegraphics[width=7cm,height=5cm]{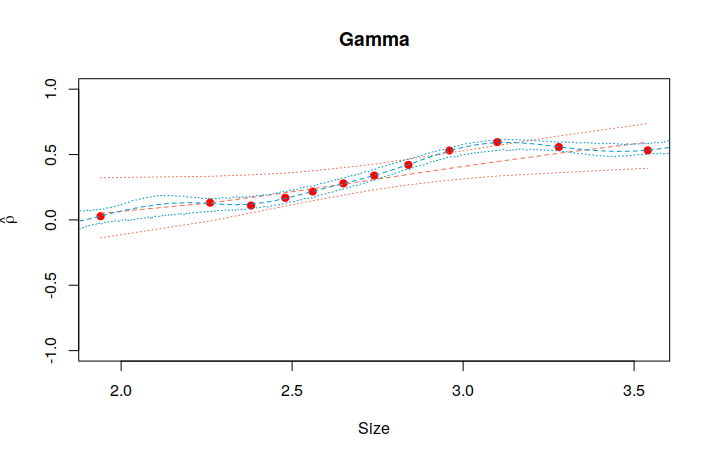}
  \caption{}
\end{subfigure}%
\begin{subfigure}{.5\textwidth}
  \centering
  \includegraphics[width=7cm,height=5cm]{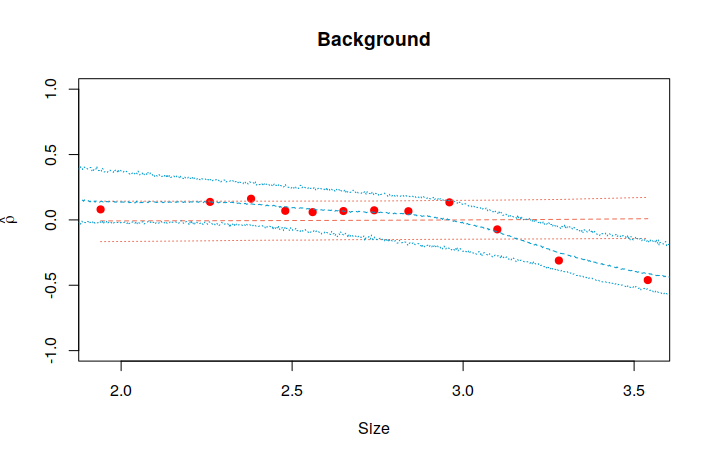}
  \caption{}
\end{subfigure}
\caption{Posterior approximations of Spearman's $\rho$ between \texttt{Length} and \texttt{M3Long} as a function of \texttt{Width} (top plots) and \texttt{Size} (bottom plots), for the observations classified as gamma rays (left plots) and background noise (right plots), as defined by \texttt{class}. The blue lines show the results obtained with the GP method, while the coral lines show the results obtained with the splines method. The inner dashed lines denote the posterior means and the dotted lines denote the 95\% credible intervals. Red dots depict the frequentist estimates of the unconditional Spearman's $\rho$ for observations belonging to the specific levels of the covariates. }
\label{fig:magic_GP}
\end{figure}

Figure \ref{fig:magic_GP} shows the approximation of the posterior mean and credible intervals of the Spearman's $\rho$ between \texttt{Length} and \texttt{M3Long} with respect to \texttt{Width} and \texttt{Size}, split by \texttt{class}, obtained with the GP and the splines methods. 
The blue lines show the results obtained with the GP method, while the coral lines show the results obtained with the splines method. The inner dashed lines denote the posterior means and the dotted lines denote the 95\% credible intervals.
Red dots depict the frequentist estimates of the unconditional Spearman's $\rho$ for observations belonging to the specific levels of the covariates. 
Similarly to the example in Section \ref{sub:energy}, splines tend to excessively smooth out the relationship between the dependence and the covariates, while the GPs better follow the data. Figure \ref{fig:magic_GP} also shows that the dependence structure among the recorded images measurements shows different patterns for gamma rays and background noise and can be used for discriminating between these two data classes. 

\begin{figure}
\centering
\begin{subfigure}{.5\textwidth}
  \centering
  \includegraphics[width=7cm,height=5cm]{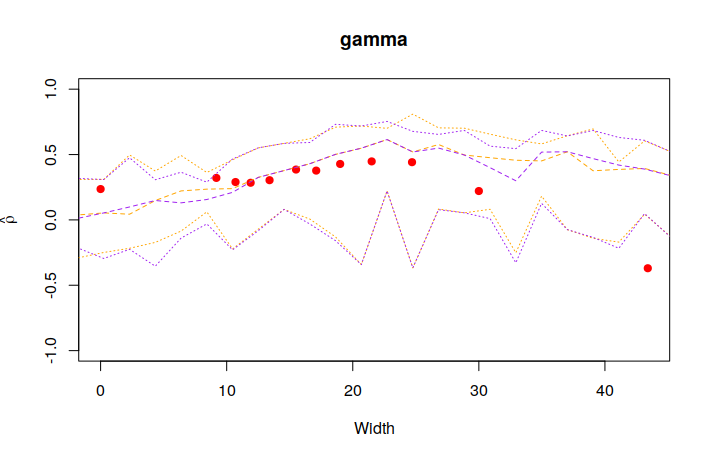}
  \caption{}
\end{subfigure}%
\begin{subfigure}{.5\textwidth}
  \centering
  \includegraphics[width=7cm,height=5cm]{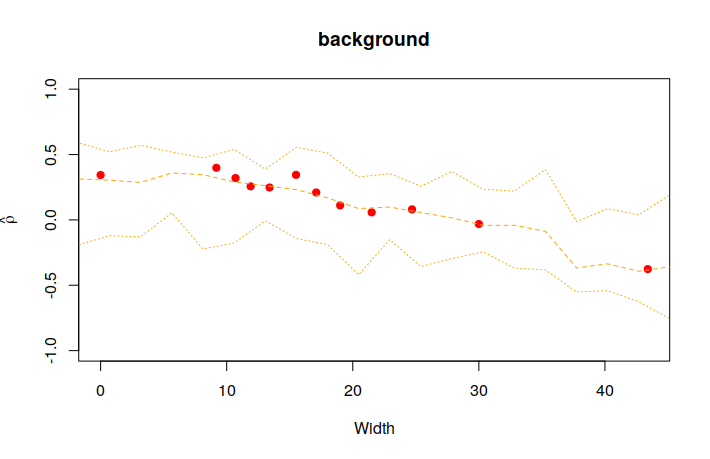}
  \caption{}
\end{subfigure}
\begin{subfigure}{.5\textwidth}
  \centering
  \includegraphics[width=7cm,height=5cm]{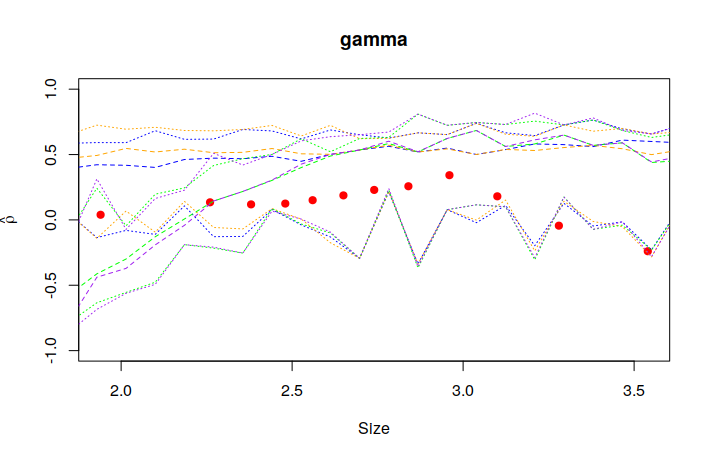}
  \caption{}
\end{subfigure}%
\begin{subfigure}{.5\textwidth}
  \centering
  \includegraphics[width=7cm,height=5cm]{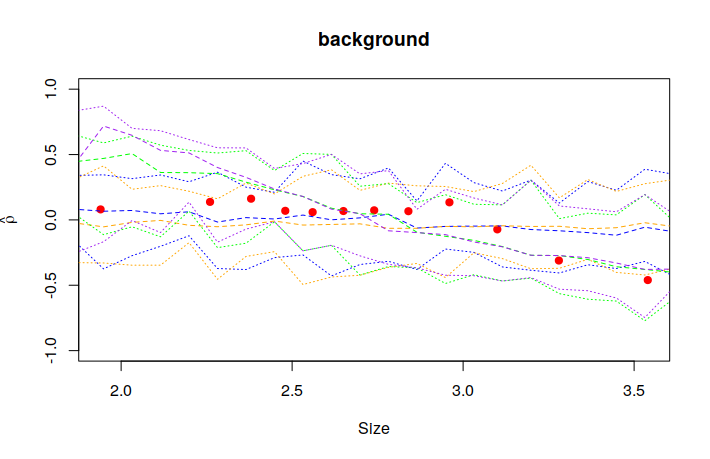}
  \caption{}
\end{subfigure}
\caption{Posterior approximations of Spearman's $\rho$ between \texttt{Length} and \texttt{M3Long} as a function of \texttt{Width} (top plots) and \texttt{Size} (bottom plots), for the observations classified as gamma rays (left plots) and background noise (right plots), as defined by \texttt{class}.
The lines show the results obtained with the methods based on the EL using: NW weights with triweigth kernel (blue), NW weights with Gaussian kernel (orange), LL weights with triweight kernel (green), LL weights with Gaussian kernel (purple). 
The inner dashed lines denote the posterior means
and the dotted lines denote the 95\% credible intervals. 
Red dots depict the frequentist estimates of the unconditional Spearman's $\rho$ for observations belonging to the specific levels of the covariate.}
\label{fig:magic_EL}
\end{figure}

Figure \ref{fig:magic_EL} shows the posterior approximations of Spearman's $\rho$ between \texttt{Length} and \texttt{M3Long} with respect to \texttt{Width} and \texttt{Size}, split by \texttt{class}, calculated with the method based on the EL, with the inconsistent estimator of the copula function using: NW weights with triweigth kernel (blue), NW weights with Gaussian kernel (orange), LL weights with triweight kernel (green), LL weights with Gaussian kernel (purple). 
The inner dashed lines denote the posterior means
and the dotted lines denote the 95\% credible intervals.
Red dots depict the frequentist estimates of the unconditional Spearman's $\rho$ for observations belonging to the specific levels of the covariate.
Similarly to the previous examples, the approximation strongly depends on the definition of the weights and the kernel functions within the weights, and the uncertainty associated with the estimates is larger than that obtained with methods based on splines or GP. Moreover, the computational cost is large, where for some approximations most of the weights describing the EL associated with different values of the functional are zero or close to zero and it is not possible to obtain accurate approximations (this is the reason why some of the estimates are not shown in the plots). In general, non linear functions seem to be less well approximated than linear functions, especially if some areas of the covariate spaces are less represented.

%
%
%
%
%
 
\section{Conclusions}
\label{sec:conclu}

In this work, we have analysed three main methodologies to approximate the posterior distribution of functionals of the dependence: Gaussian processes, methods based on the empirical likelihood, and methods based on Bayesian splines. 

We have compared the methods in terms of approximation error and precision of the estimates. 

The main advantage of all these methods is that they avoid the selection of the copula family. We have shown in practical examples that the selection of the copula function is not an easy task. In particular, when the functional of the dependence is influenced by covariates, two main difficulties arise: the number of observations for each level of the covariates can be too limited to properly select the model and the structure of the dependence can in practice change with the level of the covariate. 

Non-linear estimation procedures (like Gaussian processes and Bayesian splines) benefit from being
flexible enough to 
adequately fit the relationship between the dependence and covariates, however they need several observations for each level of the covariates to define a noisy version of the functionals to be estimated.
Such requirement can be limiting in applied contexts either because there could be only one observation for each level or because the covariate is continuous. 
In the latter case, groups of covariate values can be combined into discrete levels.
In any case, these methods can be implemented with a limited number of covariates. 

Methods based on the empirical likelihood, despite not needing replications for each covariate level, on the other hand show higher approximation errors. When using an inconsistent estimator of the copula function, the approximation seems to strongly depend on the choice of the weights and the approximation error is larger than GP- or splines-based methods. On the other hand, when using a linearised model of the functional through a Taylor's expansion, the uncertanty increases so that inference is not meaningful.

\bibliographystyle{imsart-nameyear}
\bibliography{biblio_copula.bib}

\end{document}